\documentclass[11pt,a4paper]{article}
\pdfoutput=1
\usepackage{jcappub}  
\usepackage{verbatim}
\usepackage{multirow}
\usepackage[utf8]{inputenc}
\usepackage[T1]{fontenc}
\usepackage{subcaption}
\usepackage{amsmath,amssymb}
\usepackage{physics}
\usepackage{graphicx}
\usepackage{hyperref}
\usepackage{geometry}
\usepackage{algorithm}
\usepackage{algpseudocode}
\usepackage{booktabs}
\usepackage{float} 
\usepackage{booktabs} 
\usepackage{caption}
\usepackage{xcolor}
\usepackage{soul}

\begin{document}

\title{A general criterion to compare metric and Palatini formulations with arbitrary non-minimal couplings: the case of Natural Inflation}
\author[1]{Andres Felipe Suarez Reyes,}\emailAdd{asuarezre@unal.edu.co}
\author[1]{and Juan P. Beltrán Almeida}
\emailAdd{jubeltrana@unal.edu.co}
 \affiliation[1]{Universidad Nacional de Colombia, Facultad de Ciencias, Departamento de F\'isica,\\ Av. Cra 30 \# 45-03, Bogot\'a, Colombia}

\abstract{ 
We study Natural Inflation with several types of non-minimal couplings to gravity in  metric and Palatini formalisms, focusing on how the choice of gravitational formulation affects inflationary observables. Starting from a Jordan-frame action with a general coupling function $\gamma^2(\phi)$, we examine several representative cases: periodic couplings ($\sin$ and $\cos$) and power-law couplings ($\phi^n$ with $n = 1, 2, -2$). To track  differences between the two formalisms, we employ  a diagnosis parameter $J$
defined as ${J}\equiv K_{\rm M}/K_{\rm P} -1$, where $K_{\rm M}$ and $K_{\rm P}$ are the kinetic functions in the Einstein frame in both formalisms. This parameter directly quantifies the relative enhancement of the kinetic sector in the metric formulation. For the aforementioned couplings, we compute predictions for the spectral index $n_s$ and tensor-to-scalar ratio $r$, compare them against Planck 2018 and ACT DR6 observational constraints in both formalisms. Extending the particular cases mentioned before, we derive a general classification of arbitrary coupling functions for small-field inflation depending on the behavior of $J$ at reference values during inflation: Case I with vanishing ${J}$, 
so, both formalisms become nearly degenerate; and Case II with non-vanishing ${J}$  
for which residual differences persist when evaluated at field values consistent with inflationary expansion. We find that periodic couplings are disfavored by data, whereas power-law couplings can achieve agreement for specific parameter ranges. Notably, in all the cases we found that Natural Inflation consistent with slow-roll evolution 
has $J$ either vanishing or strongly suppressed; consequently, switching the gravitational formulation yields no improvement for its observational viability.  

}

\date{\today}

\maketitle





\section{Introduction}
\label{intro}
Recent data from observational collaborations, particularly the Atacama Cosmology Telescope (ACT) \cite{ACT1,ACT2}, strongly support cosmic inflation as the driver of the Universe's early accelerated expansion. In addition to solving the initial-condition fine-tuning problems, inflation generates quantum fluctuations in the inflaton field that seed all the structures in the universe. Among various inflationary scenarios, Natural Inflation \cite{Freese:1990rb} offers a compelling, particle-physics-motivated framework driven by an axion, a pseudo-scalar Nambu-Goldstone boson, arising from the spontaneous breaking of a global shift symmetry at a scale $f$. Protected by a periodic shift symmetry via the potential $V(\phi)= \Lambda^4 (1 + \cos{(\phi/f)})$ this model remains stable against radiative corrections and possesses attractive UV completions including gravity. Furthermore, axions and axion-like particles (ALPs) naturally emerge in many places in high-energy physics and serve as viable dark matter candidates due to their suppressed interactions. Despite its many virtues, natural inflation in its simplest version is plainly excluded by current observational data \cite{Akrami2020}.

Motivated by the appealing features of Natural Inflation, we revisit the model in the presence of non-minimal couplings to gravity. Extensive literature \cite{Futamase:1987ua, Salopek:1988qh, Fakir:1990eg, Amendola:1990nn, Kaiser:1994vs, Bezrukov:2007ep, Bauer2008, Park:2008hz, Linde:2011nh, Nozari:2015ada, Kallosh:2013maa, Chiba:2014sva, Boubekeur:2015xza, Pieroni:2015cma, Salvio:2017xul, Faraoni:1996, Lucchin:1986, Abbott:1981} demonstrates that non-minimal couplings can bring simple and well motivated single-field inflation models into agreement with observational data \cite{Boubekeur:2015xza, Tenkanen:2017jih}. Non-minimally coupled Natural Inflation specifically has also been studied in several contexts (see, e.g., \cite{Nozari:2015ada, Domcke:2017fix, Ferreira_2018, BeltranAlmeida:2018ycz, dos_Santos_2024}). To broadly explore possible interactions, we examine two coupling types: periodic couplings that respect the symmetry of the inflaton potential (a cosine coupling was analyzed in \cite{Ferreira_2018, Salvio:2023cry}), and power-law couplings of the form $\alpha (\phi/M_{p})^{n}$, where $\alpha$ is a dimensionless parameter.


It is also well known that non-minimal gravitational couplings lead to distinct dynamics in the connection degrees of freedom depending on the chosen formalism. In the Palatini formalism, for instance, treating the metric and the connection as independent fields alters the transition from the Jordan to the Einstein frame, significantly affecting inflationary predictions \cite{Bauer2008,Racioppi:2017,Jarv:2020,Jinno:2020,Tenkanen:2020review,Gialamas_2023,Dioguardi:2026}. While these discrepancies are most pronounced in large-field inflation models, this work presents a detailed comparative analysis of both formalisms focusing on natural inflation and small-field models. Extending previous studies of natural inflation with non-minimal gravitational couplings in the Palatini formulation \cite{Bostan_2023_BICEP_LINEAL,Bostan:2025zdt,Antoniadis:2019NaturalPalatini}, our work focuses and complement the discussion on three main aspects: a) We study general periodic couplings and we do a general comparison of metric and Palatini and we employ a  criterion to distinguish when both formalisms exhibit noticeable quantitative differences. b) We explore general power law couplings including odd and negative exponents 
and classify non-minimal coupling functions for small-field models where the two formalisms diverge significantly.
c) We discuss universality of $\alpha$-attractors \cite{Galante_2015,Racioppi:2018zoy} 
demonstrating that the attractor behavior encoded in the pole structure is independent of the formalism, whereas the definition of the residue is formalism-dependent.

This paper is organized as follows: in Section~\ref{MP_review} we review the essential
differences between the metric and Palatini formalisms and frame it in slow roll inflationary dynamics.
In section \ref{sec:P_factor} we discuss the Palatini
Factor $\mathcal{P}$ and the deviation from unity parameter $J$ as a measure of the relative enhancement of the kinetic sector
between the two formulations and derive a model-independent classification of
couplings into two classes according to the behavior of the Palatini Factor in the
small-field regime.  In
Section~\ref{sec:Natural_periodic} we specialize to Natural Inflation and study several representative coupling functions --- cosine, sine, and power law --- comparing
the resulting predictions for $n_s$ and $r$ in both formalisms against the Planck 2018
and ACT DR6 constraints. Section~\ref{sec:alpha_attractors} briefly discusses the connection
between this framework and $\alpha$-attractors, showing that the metric--Palatini
degeneracy in that context maintains the pole structure but redefines by a constant term the residue of the kinetic term. 
In Section~\ref{sec:general_results} we generalize the previous results to arbitrary
periodic and power-law couplings. Finally, Section~\ref{sec:conclusions} summarizes our conclusions
and discusses possible extensions of this work; details of the numerical implementation
are collected in Appendix~\ref{app:numerical}.

\section{Brief Review of Metric and Palatini Formalisms and inflation}
\label{MP_review}
In this section, we briefly review the main differences between the Metric and the Palatini formalisms and their implications for the dynamics of the system. 
The essential difference between both formalisms lies in the choice of independent variables used in the variational principle. In the metric formalism the action is varied only with respect to the metric, assuming the connection to be the Levi--Civita connection. On the other hand, in the Palatini formalism the metric and the connection are treated as independent variables and the action is varied with respect to both. It is worth noting that both formalisms lead to identical equations of motion in General Relativity (GR). However, when modified gravity theories are considered, 
the two approaches generally lead to different dynamics.  For instance, in the framework of $f(R)$ gravity one obtains the following field equations for the two formalisms \cite{Sotiriou2010}:

\paragraph{Metric Formalism}

\begin{equation*}
f'(R)R_{\mu\nu}-\frac{1}{2}f(R)g_{\mu\nu}
-\left(\nabla_\mu\nabla_\nu+g_{\mu\nu}\Box\right)f'(R)
=\kappa T_{\mu\nu}.
\end{equation*}

\paragraph{Palatini Formalism}

\begin{equation*}
f'(R)R_{\mu\nu}-\frac{1}{2}f(R)g_{\mu\nu}
=\kappa T_{\mu\nu},
\end{equation*}

\begin{equation*}
\nabla_\lambda\left(\sqrt{-g}\,f'(R)g^{\mu\nu}\right)=0.
\end{equation*}
 \noindent
This is also true when non-minimal couplings between matter degrees of freedom and gravity are taken into account. This is the case that we are going to study in this work.


\subsection{Transformation to the Einstein Frame}
As mentioned before, when non-minimal couplings to gravity are involved, differences between the two formalisms arise. With this in mind, we perform the usual conformal transformation of the metric in order to move from the Jordan frame to the Einstein frame. The initial action in the Jordan frame for these models is:
\begin{equation}\label{S_Jordan}
S_J = \int d^4x \sqrt{-\tilde g}\left[\frac{M_{pl}^2}{2}\,\gamma^2(\phi)\,\tilde R
- \frac{1}{2}\,\tilde g^{\mu\nu}\partial_\mu\phi\,\partial_\nu\phi - V(\phi)\right],
\end{equation}
\\
where $\gamma^2(\phi)$ is the non minimal coupling function.  After performing the  conformal transformation $g_{\mu\nu} = \gamma^2 \tilde{g}_{\mu\nu}$ and rescaling the potential, the action in the Einstein frame takes the form
\begin{equation}
S_E=\int d^4x\sqrt{- g}
\left[
\frac{M_{\rm {pl}}^2}{2} R
-\frac{1}{2}K(\phi) g^{\mu\nu}\partial_\mu\phi\partial_\nu\phi
-U(\phi)
\right].
\label{eq:Eaction}
\end{equation}
where $R$ is the Ricci scalar associated with the metric $ g_{\mu\nu}$, $K(\phi)$ is the kinetic factor in which the differences between formalisms are encoded, and $U(\phi)$ is the potential in the Einstein frame. Their formal expressions are: 
\begin{equation}
U(\phi)=\frac{V(\phi)}{\gamma(\phi)^4}, \qquad \text{and} \qquad K(\phi)=\frac{1+6{\rm p}M_{\rm {pl}}^2(\gamma_{,\phi})^2}{\gamma^2},
\end{equation}  
in which p takes the values ${\rm p} = 1, 0$ for the metric and Palatini formulations respectively. In this work we denote the metric kinetic  factor as $K_{\rm M}(\phi)$, the Palatini kinetic factor as $K_{\rm P}(\phi)$ and for convenience of our analysis we define the \textbf{Palatini Factor} ($\mathcal{P}$) as:
\begin{equation} 
    \mathcal{P} \equiv \frac{K_{\rm M}}{K_{\rm P}}.
    \label{Palatini_factor}
\end{equation}

\subsection{Canonical Field}

For both formalisms, the canonical field is defined as
\begin{equation}
\frac{d\chi}{d\phi}=\sqrt{K(\phi)}.
\label{fieldx}
\end{equation}
Using the definition \eqref{Palatini_factor} we can write the relation between the canonical fields in the two formalisms as
\begin{equation}
\frac{d\chi_{\rm P}}{d\phi}
=\sqrt{K_{\rm P}(\phi)}
=\sqrt{\frac{K_{\rm M}(\phi)}{\mathcal{P}(\phi)}},
\end{equation}
which implies that the relation between canonical fields in both formalism is: 
\begin{equation}
\frac{d\chi_{\rm P}}{d\phi}
=
\frac{1}{\sqrt{\mathcal{P}(\phi)}}
\frac{d\chi_{\rm M}}{d\phi}.
\end{equation}
In both cases, after canonicalization, the action \eqref{eq:Eaction} takes the form:
\begin{equation}\label{action_canonical}
S_E=\int d^4x\sqrt{- g}
\left[
\frac{M_{\rm {pl}}^2}{2} R
-\frac{1}{2}  g^{\mu\nu}\partial_\mu\chi\partial_\nu\chi
-U(\chi)
\right],
\end{equation}
and we can follow the standard canonical inflationary dynamics solving equations of motion in minimally coupled general relativity with the respective canonical field $\chi$ in each formalism.

\subsection{Inflationary Dynamics}
\label{sec:dyna}
In terms of the canonical field $\chi$, the potential in the Einstein frame is given by
\begin{equation}
U(\chi)=\frac{V(\phi(\chi))}{\gamma(\phi(\chi))^4},
\end{equation}
in which we implicitly use the relation \eqref{fieldx} defining $\chi$. 
The background dynamics of the inflaton field, derived from the action \eqref{action_canonical} are governed by the usual Friedmann equations 
\begin{equation}
3M_{\text{pl}}^{2}H^{2}=\frac{1}{2}\dot{\chi}^{2}+U(\chi),\quad  \dot{H}=-\frac{\dot{\chi}^{2}}{2M_{\rm {pl}}^{2}},
\end{equation}
together with the Klein–Gordon equation for the scalar field:
\begin{equation}
\ddot{\chi}+3H\dot{\chi}+U_{,\chi}(\chi)=0,
\end{equation}
where $U_{,\chi}$ denotes derivative with respect to $\chi$.   Inflation occurs when the slow-roll conditions $\dot{\chi}^{2}\ll U(\chi)$ and $\ddot{\chi}\ll3H\dot{\chi}$ are satisfied. For this single field scenario, slow-roll evolution can be achieved when the slow-roll parameters for the potential are small. These parameters are defined as  usual
\begin{equation}
\epsilon = \frac{M_{\rm {pl}}^2}{2}\left(\frac{U_{,\chi}}{U}\right)^2,
\qquad
\eta = M_{\rm {pl}}^2\frac{U_{,\chi \chi}}{U},
\end{equation}
and inflation is obtained when $\epsilon\ll 1, \eta \ll 1$. 
The number of e-folds between the field values $\phi_{*}$ at horizon crossing and $\phi_{\text{end}}$ at the end of inflation is given by
\begin{equation}
N=\int_{\phi_{\text{end}}}^{\phi_*}\frac{\sqrt{K(\phi)}}{\sqrt{2\epsilon}}\frac{d\phi}{M_{\rm {pl}}}.
\end{equation}
The potential slow-roll parameters can be written
in terms of the original field $\phi$ as

\begin{equation}
\epsilon=\frac{M_{\rm {pl}}^2}{2K(\phi)}
\left(\frac{U_{,\phi}}{U}\right)^2,\quad 
\eta=\frac{M_{\rm {pl}}^2}{U} \left[\frac{U_{,\phi\phi}}{K(\phi)}-\frac{K_{,\phi}(\phi)U_{,\phi}}{2K(\phi)^2}\right].
\end{equation}
Therefore, the difference between the metric and Palatini formalisms
enters only through the kinetic factor $K(\phi)$.The inflationary observables that we consider here are the spectral index for scalar perturbations $n_s$ and the tensor-to-scalar ratio $r$, which are computed from the slow-roll parameters as 
\begin{equation}
    n_s = 1 - 6\epsilon^* + 2\eta^*, \qquad r = 16\epsilon^*,
    \label{eq:obs}
\end{equation}
evaluated at $\phi_*$, the field value at which the pivot scale 
$k_* = 0.002\,\text{Mpc}^{-1}$ crosses the horizon $N_*$ e-folds before the end of inflation.

\section{The Palatini Factor as a General Criterion}
\label{sec:P_factor}
When studying non-minimally coupled gravity in both the metric and Palatini formalisms, a natural question arises: will the two formulations remain equivalent, as they are in General Relativity, or should a genuine difference be expected? Motivated by this question, we introduce the Palatini factor, which quantifies precisely when the two formalisms are degenerate and when they are not. As already introduced in Section \ref{MP_review}, the Palatini factor is defined as
\begin{equation}
\mathcal{P}(\phi) \equiv \frac{K_{\rm M}}{K_{\rm P}}=1+6M_{\rm {pl}}^2(\gamma_{,\phi})^2.
\label{Palatini_factor_gamma}
\end{equation}
Since the additional metric contribution to the kinetic sector is positive definite, one always has $K_{\rm M}\geq K_{\rm P}$, and therefore $
\mathcal{P}\geq1$. Thus, the quantity $\mathcal{P}-1$ directly measures the relative enhancement of the kinetic sector in the metric formulation with respect to the Palatini case.\\
The comparison \textbf{criterion} is the deviation of the Palatini factor from unity: if $\mathcal{P}=1$, no differences are expected, since $K_{\rm M}=K_{\rm P}$ and the two formalisms reduce to the same Einstein-frame action. Conversely, for $\mathcal{P}>1$, differences between the formalisms are expected. It is worth emphasizing that, although such differences exist analytically, they may be physically negligible, and that $\mathcal{P}$ is not itself a physical observable: it is a comparison tool between two mutually exclusive formalisms, gravity being either metric or Palatini, but never both simultaneously.

\subsection{Inflationary Palatini Factor}

The Palatini factor enters the inflationary dynamics through the canonical field $\chi$, and becomes explicit once the relevant equations are rewritten in terms of the original field $\phi$. For our purposes, we are interested in the slow-roll parameters $\epsilon$ and $\eta$ and in the number of $e-$folds $N$. They compare as
\begin{equation}
\epsilon_{\rm P}(\phi) = \mathcal{P}(\phi)\,\epsilon_{\rm M}(\phi), \qquad
\eta_{\rm P}(\phi) = \mathcal{P}(\phi)\,\eta_{\rm M}(\phi) + \frac{M_{\rm pl}^2\,\mathcal{P}_{,\phi}}{2K_{\rm M}}\,\frac{U_{,\phi}}{U}.
\label{eq:slow_roll_relations}
\end{equation}
\begin{equation}
    N_M(\phi)=\int_{\phi_{\rm{end}}^{(M)}}^{\phi} \mathcal{P}(\phi) N_{P,\phi} \, d\phi.
    \label{eq:efolds_relation}
\end{equation}
These relations hold at any fixed field value $\phi$. In the trivial case $\mathcal{P}(\phi)=1$, they reduce to $\epsilon_{\rm P}=\epsilon_{\rm M}$ and $\eta_{\rm P}=\eta_{\rm M}$ identically; moreover, $K_{\rm M}=K_{\rm P}$ implies an identical $e$-fold integrand, so that $\phi_*^{(\rm M)}=\phi_*^{(\rm P)}$, and $n_s$, $r$ coincide exactly.

For $\mathcal{P}\not=1$, however, this simplicity is lost: since the number of $e$-folds depends explicitly on $K(\phi)$, the horizon-crossing field value itself differs in the two formalisms, $\phi_*^{(\rm M)}\neq\phi_*^{(\rm P)}$, and
\begin{equation}
\Delta r \equiv r_{\rm P}-r_{\rm M} = 16\left[\mathcal{P}\big(\phi_*^{(\rm P)}\big)\,\epsilon_{\rm M}\big(\phi_*^{(\rm P)}\big) - \epsilon_{\rm M}\big(\phi_*^{(\rm M)}\big)\right].
\end{equation}
Departures of $\mathcal{P}$ from unity signal that a difference exists, even though its precise size depends on the full slow-roll trajectory between $\phi_*^{(\rm M)}$ and $\phi_*^{(\rm P)}$, and not uniquely on $\mathcal{P}$ evaluated at a single point.
\subsection{The $J$-Deviation}
We isolate the departure from unity of the ${\cal P}$ factor by  the quantity $J\equiv \mathcal{P} -1$. This quantity coincides with the ratio $\omega$ introduced in Ref.~\cite{Karam:2021wzz} (Eq.~(5.1) therein), where it was used to delimit the field domains in which the metric contribution to the kinetic term dominates or is suppressed, with an explicit analysis for monomial couplings in the context of $\xi$-attractors. Here we use the same quantity in a different direction: we track how it propagates exactly into the observable differences $\Delta r$, $\Delta n_s$ and the horizon-crossing shift $\Delta\phi_*$, and we use its small-field behaviour to classify arbitrary coupling functions.
Since the Palatini factor $\mathcal{P}$ and $J$ carry the same information —differing only by an additive constant—, we use them interchangeably throughout the remainder of this work, choosing whichever quantity makes a given expression clearer. Then,   Eq.~\eqref{eq:slow_roll_relations} is re-expressed as
\begin{equation}
\epsilon_{\rm P}(\phi) = \epsilon_{\rm M}(\phi) + J(\phi)\,\epsilon_{\rm M}(\phi), \qquad
\eta_{\rm P}(\phi) = \eta_{\rm M}(\phi) + J(\phi)\,\eta_{\rm M}(\phi) +\Xi(\phi),
\label{eq:slow_roll_J}
\end{equation}
where $\Xi(\phi)= \frac{M_{\rm pl}^2\,J_{,\phi}(\phi)}{2K_{\rm M}(\phi)}\,\frac{U_{,\phi}}{U}$ and the number of $e$-fold satisfy the exact relation
\begin{equation}
N_{\rm M}(\phi) = \int_{\phi^{(M)}_{\rm end}}^{\phi} \big(1+J(\phi')\big)\,N_{{\rm P},\phi}(\phi')\,d\phi'.
\label{eq:efolds_J}
\end{equation}
Each expression separates explicitly into a Palatini term plus a correction controlled by $J$, vanishing identically when $J=0$. We now track how this correction propagates into the horizon-crossing shift $\Delta\phi_*\equiv\phi_*^{(\rm P)}-\phi_*^{(\rm M)}$, and from there into $\Delta r$ and $\Delta n_s$.
\subsubsection{Tracking differences in the observables with $J(\phi)$}
Applying the mean value theorem to Eq.~\eqref{eq:slow_roll_J} gives the exact differences, $\Delta r \equiv r_{\rm P}\big(\phi_*^{(\rm P)}\big)-r_{\rm M}\big(\phi_*^{(\rm M)}\big)$ and $ \Delta\eta \equiv \eta_{\rm P}\big(\phi_*^{(\rm P)}\big)-\eta_{\rm M}\big(\phi_*^{(\rm M)}\big) $ as follows
\begin{align}
\Delta r &=16 \Big[J\big(\phi_*^{(\rm P)}) \epsilon_{\rm M}\big(\phi_*^{(\rm P)}\big)+\epsilon_{\rm M_{,\phi}}(\mu)\Delta\phi_*\Big], \label{eq:delta_r_exact}\\
\Delta\eta &= J(\phi_*^{(\rm P)})\,\eta_{\rm M}(\phi_*^{(\rm P)}) + \Xi(\phi_*^{(\rm P)}) + \eta_{{\rm M},\phi}(\mu)\,\Delta\phi_*, \label{eq:delta_eta_exact}
\end{align}
Similarly, from Eq.~\eqref{eq:efolds_J} and the mean value theorem, the horizon-crossing shift can be written as
\begin{equation}
\Delta\phi_* = -\,\frac{1}{N_{,\phi}^{(\rm P)}(\mu)}\left[N_*\Big\langle \frac{J}{\mathcal{P}}\Big\rangle_N + \frac{N_{,\phi}^{(\rm P)}(\mu')}{\epsilon_{{\rm P},\phi}(\mu)}\,J(\phi_{\rm end}^{(M)})\right], \label{eq:delta_phi_exact} \qquad \langle F\rangle_N \equiv \frac{1}{N_*}\int_0^{N_*} F\, dN_{\rm M},
\end{equation}
\textbf{where $\mu$} denotes, in each equation, some point between $\phi_*^{(\rm M)}$ and $\phi_*^{(\rm P)}$ guaranteed by the mean value theorem. In the expressions above we abuse notation and use the same symbol $\mu$ throughout, though its value generally differs from one expression to another. Using $n_s=1-6\epsilon+2\eta$, we get
\begin{equation}
\Delta n_s \equiv n_{s,{\rm P}}-n_{s,{\rm M}} = -\frac{3}{8}\,\Delta r + 2\,\Delta\eta.
\label{eq:delta_ns_exact}
\end{equation}
Equations~\eqref{eq:delta_r_exact}, \eqref{eq:delta_eta_exact} and \eqref{eq:delta_phi_exact} put the qualitative criterion of ~\cite{Karam:2021wzz} on a quantitative footing. Every term on their right-hand sides is proportional to $J$, evaluated at horizon crossing ($J_*$), at the end of inflation ($J(\phi^M_{\rm end})$), or averaged along the trajectory ($\langle J/{\cal P}\rangle_N$), and each one is further multiplied by a slow-roll quantity. In particular, at  order  $J_{\rm max}\ll 1$, Eq.~\eqref{eq:delta_phi_exact} gives
\begin{equation}
   \Delta\phi_* \approx -\,\frac{N_*\langle J\rangle_N}{N_{,\phi}(\phi_*)}, \label{eq:delta_phi_aprox} 
\end{equation}
since $\langle J\rangle_N \leq J_{\max}$, it follows that $\Delta\phi_* \ll 1$, and no differences between the two formalisms are expected. Taking into account the slow-roll conditions, the order of the observables can be estimated as\footnote{This uses $K_{\rm M} = \mathcal{O}(1)$ and $\eta_{\rm M} \sim \epsilon_{\rm M}$, together with the assumption that $J_{,\phi}(\phi_*) = \mathcal{O}(J)$ is not anomalously large.}
\begin{equation}
    \Delta r \approx \mathcal{O}(\epsilon_{\rm M}\, J), \label{eq:delta_orden} \qquad
    \Delta n_s \approx \mathcal{O}(\sqrt{\epsilon_{\rm M}}\, J), 
\end{equation}
\subsection{General Coupling Functions}

Consider a small-field \footnote{Throughout this work, ``small-field'' refers to horizon crossing occurring close to the maximum of the potential} inflationary model with an arbitrary, well-behaved analytic coupling $g(\phi;\theta)\equiv\gamma^2(\phi;\theta)$, which may additionally depend on other model parameters, collectively denoted $\theta$ (such as a decay constant $f$).Writing $\gamma = \sqrt{g}$, so that $\gamma_{,\phi} = g_{,\phi}/(2\sqrt{g})$
\begin{equation}
J(\phi;\theta) = \frac{3M_{\rm pl}^2}{2}\,\frac{g_{,\phi}(\phi;\theta)^2}{g(\phi;\theta)},
\label{eq:J_general}
\end{equation}
where all derivatives are taken with respect to $\phi$ at fixed $\theta$.
which, at leading order in $\phi_*\ll1$, gives
\begin{equation}
J_* \simeq \frac{3M_{\rm pl}^2}{2}\,\frac{g_{,\phi}(0;\theta)^2}{g(0;\theta)}.
\label{eq:J_leading}
\end{equation}
As discussed above, the relevant classification concerns the order of $J_*$: for $J_* \ll 1$, no observable differences are expected.. This gives rise to two natural classes:
 \paragraph{Case I: $J_*\ll1$. \textbf{Vanishing J}}. As shown in Eqs.~\eqref{eq:delta_orden},  no differences in the observables are expected. If $g$ is well-behaved, the condition $g_{,\phi}(0;\theta)=0$ serves as a flag for near-degeneracy between the two formalisms, since {we should do a second order expansion, so}
\begin{equation}
J_* \simeq \frac{3M_{\rm pl}^2}{2}\,\frac{g_{,\phi\phi}(0;\theta)^2}{g(0;\theta)}\,\phi_*^2,
\end{equation}
{where $\phi_{*}$ is small in this case. Obviously, in large field models, this behavior is responsible for large differences}. One caveat remains, however: if the ratio $\frac{g_{,\phi\phi}(0;\theta)^2}{g(0;\theta)}$ is anomalously large, near-degeneracy may not hold. {This can happen if derivatives of $g$ near zero are not well behaved or when further dependence of the model parameters $\theta$ remain and this turns out to be large. Below we discuss a specific example.} 

\paragraph{Case II: Non-vanishing J.} Here, the first derivative serves as a flag, now with a non-vanishing value, $g_{,\phi}(0;\theta)\neq0$. One caveat remains: the residual value of $J_*$ may be further suppressed or enhanced through $\theta$, rather than through $\phi_*$. Consequently, near-degeneracy is not completely guaranteed within case II.

\section{Natural Inflation with Non-minimal Couplings}
\label{sec:Natural_periodic}
Now we focus on the specific case of {\it Natural Inflation}, then we consider the action \eqref{S_Jordan} with the potential 


\begin{equation}
V(\phi)=\Lambda^4
\left[1+\cos\left(\frac{\phi}{f}\right)\right].
\end{equation}
In order to ensure a well-defined gravitational sector, the coupling function $\gamma^2(\phi)$ must satisfy the condition
$ \gamma^2(\phi) > 0,$
 so that the effective Planck mass $M_{\text{eff}}^2 = M_{\text{pl}}^2 \gamma^2(\phi)$ remains positive. 
This requirement guarantees the correct sign of the Einstein--Hilbert term and avoids ghost-like instabilities in the gravitational dynamics.
\subsection{Inflationary Regime for Different Couplings}
\label{4.1}
For our purpose, we explore different functional forms of $\gamma(\phi)^2$ in the inflationary regime to assess their impact on the effective dynamics and observable predictions. For single field models, we use a single non minimal coupling parameter $\alpha$. It is worth noting that most of the values of the non minimal coupling parameter \(\alpha\) considered for each \(\gamma\) are chosen based on inflationary requirements, namely the presence of a sufficiently flat potential, or suppressed slow-roll potential parameters, together with numerical exploration. For each case, we show the comparison with the $n_s-r$ contours from Planck 2018 \cite{Akrami2020} and 
ACT DR6 \cite{ACT2}. 

\subsubsection{Coupling A: $\gamma^2(\phi)=1+\alpha\left(1+\cos(\frac{\phi}{f})\right)$}
\begin{figure}[H]
    \centering
        \includegraphics[width=0.6\textwidth]{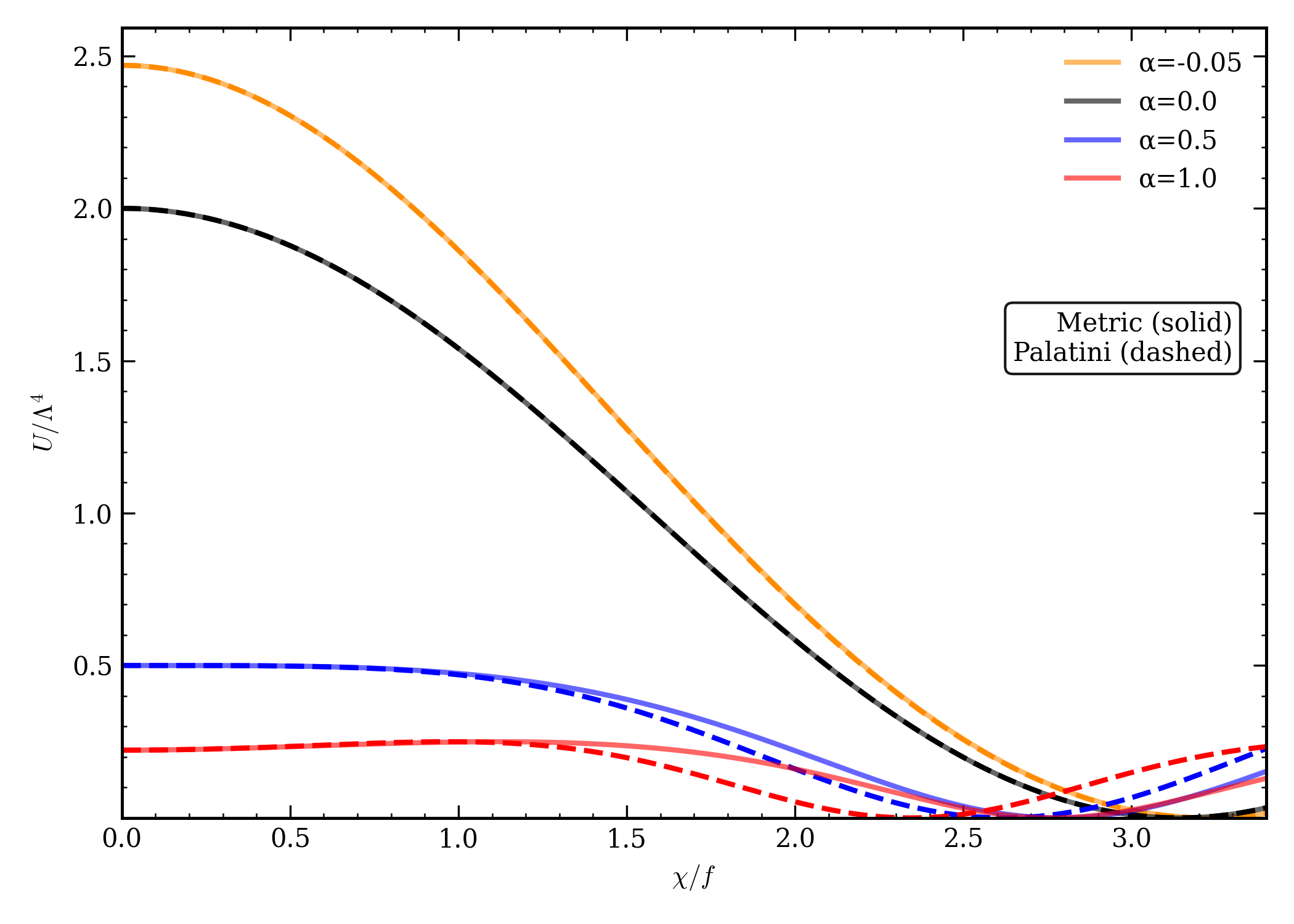}
    \caption{Potential in the Einstein frame $U/\Lambda^4$ as a function of $\chi/f$ for different values of $\alpha$ ($f=1$). For cosine non-minimal coupling, superposition of both formalisms. As $\alpha$ increases, the potential becomes progressively flatter.}
    \label{fig:potential_comparisonA}
\end{figure}

This coupling preserves both the $U(1)$ shift symmetry  and the CP symmetry $\mathbb{Z}_2$ of the original Natural Inflation potential. 
This type of coupling function was studied in \cite{Ferreira_2018, Salvio:2023cry} and for the sake of comparison, we follow the notation developed in these references. We solve numerically Eq.~\eqref{fieldx} and the slow-roll parameters for both formalisms.
In this model analytic expressions for the slow-roll parameters 
can also be obtained. 
This analytic solution is used to calibrate the numerical results. 
The largest absolute error between the analytic and the numerical evaluation of $\epsilon$ analytic  is of order $10^{-5}$, while within the inflationary regime
the difference between the numerical and analytic solutions is typically
$10^{-19}$, $10^{-11}$. These small discrepancies confirm that the numerical solution provides an accurate description of the dynamics. The potential in the Einstein frame has the form shown in Fig.~\ref{fig:potential_comparisonA}.

\begin{figure}[H]
    \centering
    \begin{subfigure}[t]{0.485\textwidth}
        \centering
        \includegraphics[width=\linewidth]{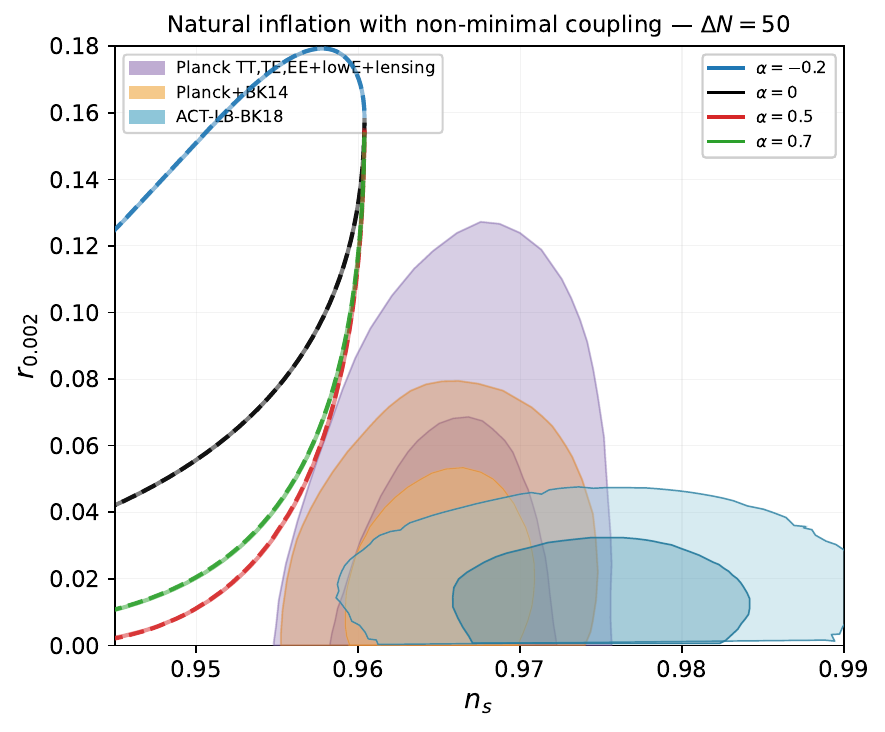}
        \label{fig:nsr_N50}
    \end{subfigure}
    \hfill
    \begin{subfigure}[t]{0.485\textwidth}
        \centering
        \includegraphics[width=\linewidth]{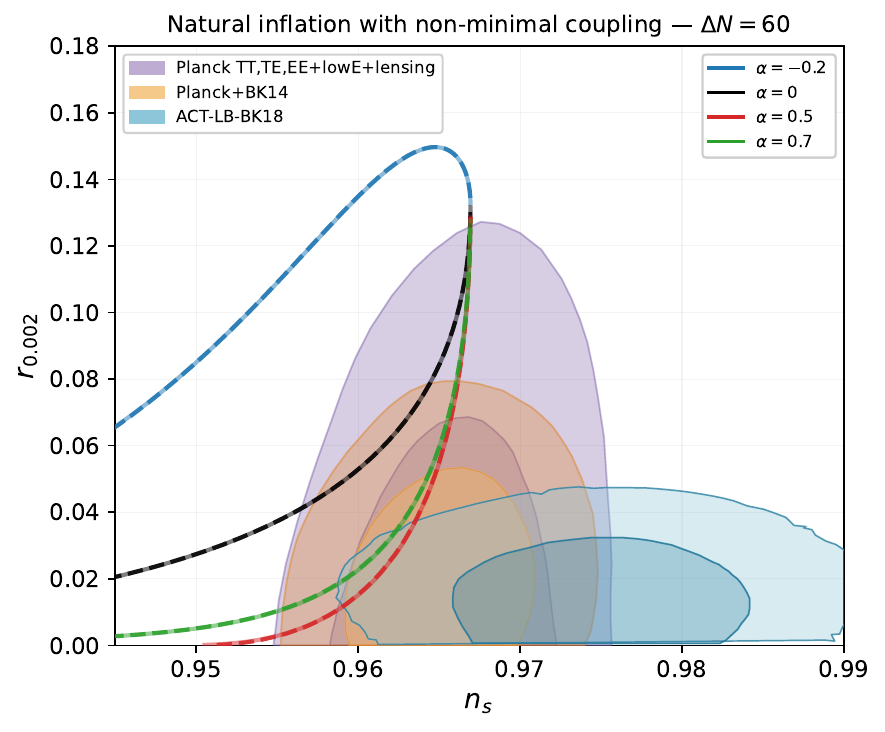}
        \label{fig:nsr_N60}
    \end{subfigure}  
    \caption{Predictions for $n_s$ and $r$ for the cosine non-minimal coupling. 
Solid lines correspond to the metric formalism and dashed lines to the 
Palatini formalism, for $N=50$ and $N=60$ with $3 < f < 80$ 
(in units of $M_{\mathrm{pl}}$). Shaded regions show the 68\% and 95\% confidence levels from Planck 2018 \cite{Akrami2020}\protect\footnotemark and 
ACT DR6 \cite{ACT2}.}
    \label{fig:nsr_coseno}
\end{figure}

\noindent 
The slow-roll parameter $\epsilon$ acquires the form:
\begin{equation}
\epsilon(\phi)_{\rm M,P}
=
\frac{M_{\rm pl}^{2}}{2f^{2}K(\phi)_{\rm M,P}}
\frac{\tan^2\!\left(\frac{\phi}{2f}\right)
\left[
1-\alpha\left(1+\cos\!\left(\frac{\phi}{f}\right)\right)
\right]^2
}{
\left[
1+\alpha\left(1+\cos\!\left(\frac{\phi}{f}\right)\right)
\right]^2},
\end{equation}
where $K_{\rm M,P}$ is the kinetic factor in metric and Palatini formalism.  
The condition $\gamma^2>0$ imposes the restriction $\alpha>-\frac{1}{2}$. The Palatini factor is:
\begin{equation}
    \mathcal{P}(\phi) = 1 + \frac{3M_{\rm pl}^2\,\alpha^2\sin^2(\phi/f)}{2f^2\,\gamma^2(\phi)}.
\end{equation}
During the inflationary regime the field is close to the maximum of the potential,
so that $\phi/f \ll 1$. Therefore one can use the small-angle approximation and obtain:
\begin{equation}
    \mathcal{P} \simeq 1+\frac{3M_{\rm pl}^2\,\alpha^2}{2f^4(1+2\alpha)}\,\phi^2 
\end{equation}
Evidence that the small-field approximation is adequate is provided by the fact that $\gamma^2\simeq1+2\alpha$, $K_{\rm M}\simeq K_{\rm P}$, and the resulting canonical field $\chi\simeq\phi/\sqrt{1+2\alpha}$  reproduce the numerical value of $\chi(\phi_*)$ with a relative error of about $0.3\%$ for both formalisms, with parameters $\alpha=0.5$ and $f=1$.
The observable predictions in the $(n_s,r)$ plane are shown in Fig. ~\ref{fig:nsr_coseno}. 
\footnotetext{Contour data for Planck and Planck BK14 extracted from \url{https://github.com/kdemirel/Planck-constraints-r-vs-ns-plot}.}

As in the case of minimally coupled natural inflation, this coupling reaches the observationally favored (shaded) region for large values of $f$, which corresponds to $\mathcal{P}\to1$ (equivalently $J\to0$); the two formalisms therefore become nearly degenerate in this limit. 
\subsubsection{Coupling B: $ \gamma^2(\phi)=1+\alpha\sin(\frac{\phi}{f})$}
\begin{figure}[H]
    \centering
    \includegraphics[width=0.6\linewidth]{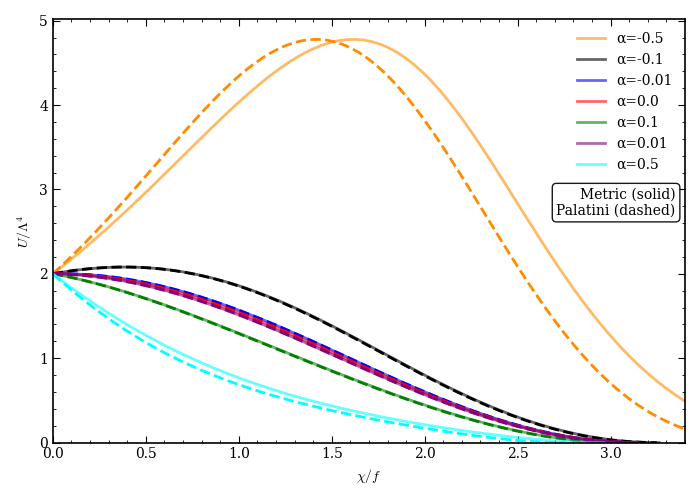}
    \caption{Potential in the Einstein frame $U/\Lambda^4$ as a function of $\chi/f$ for different values of $\alpha$ ($f=1$).}
    \label{fig:potential_comparisonB}
\end{figure}

\begin{figure}[H]
    \centering
    \begin{subfigure}[t]{0.496\textwidth}
        \centering
        \includegraphics[width=\linewidth]{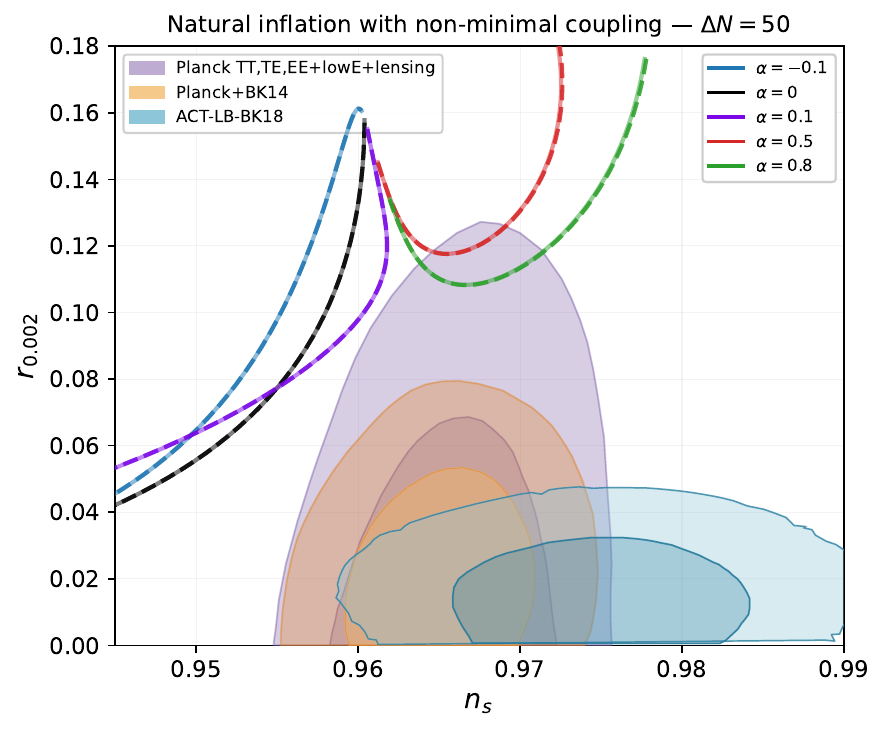}
    \end{subfigure}
    \hfill
    \begin{subfigure}[t]{0.496\textwidth}
        \centering
        \includegraphics[width=\linewidth]{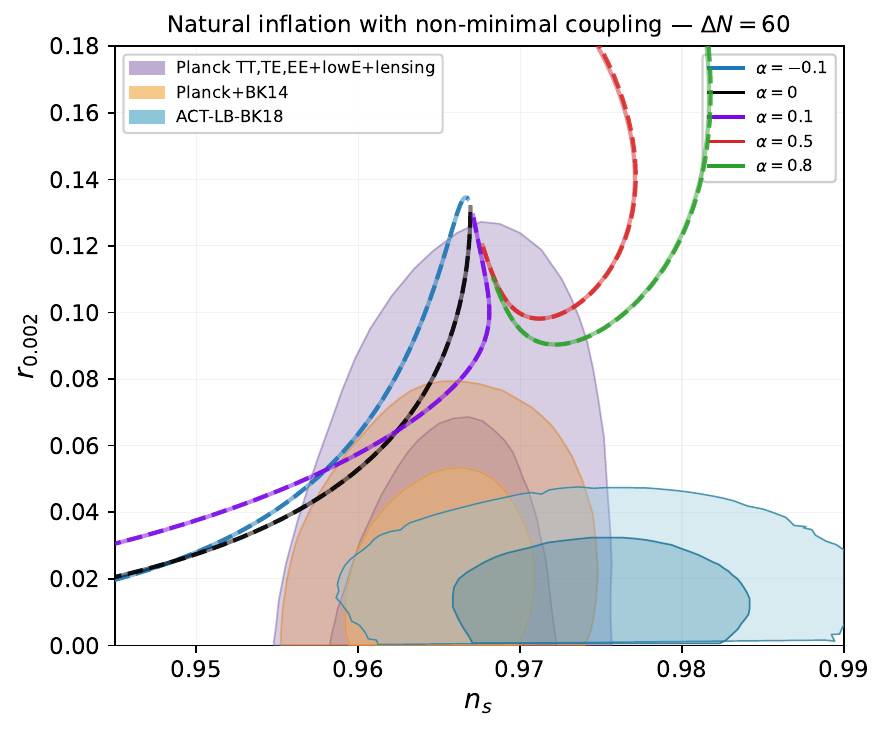}   
    \end{subfigure}
    \caption{Predictions for $n_s$ and $r$ for the sine non-minimal 
coupling. Solid and dashed lines correspond to the 
metric and Palatini formalisms with $3 < f < 80$ 
(in units of $M_{\mathrm{pl}}$). Shaded regions as in Fig.~\ref{fig:nsr_coseno}. }
    \label{fig:nsr_seno}
\end{figure}

\begin{figure}[H]
    \centering
    \includegraphics[width=0.5\linewidth]{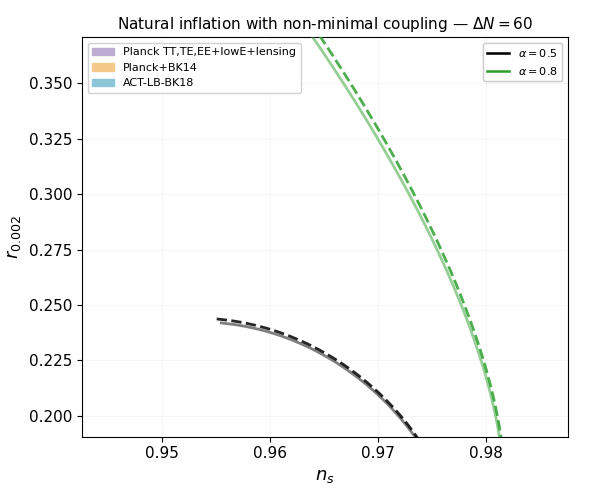}
    \caption{{Predictions for $n_s$ and $r$ for the sine non-minimal 
coupling. Solid and dashed lines correspond to the 
metric and Palatini formalisms with $2< f < 6$ 
(in units of $M_{\mathrm{pl}}$). These values are far above the confidence bands. }}
    \label{fig:pala_sin_small}
\end{figure}

This coupling preserves the shift symmetry $U(1)$ and violates the CP symmetry $\mathbb{Z}_2$ of the original Natural Inflation potential. It should be noted, however, that CP symmetry is not a fundamental constraint of the model but rather an accidental feature of the cosine potential \cite{Freese:1990rb}. In this case the potential in the Einstein frame has the form, shown in Fig. \ref{fig:potential_comparisonB}.
\noindent
The slow-roll parameter $\epsilon$ and the Palatini factor become respectively:
\begin{equation}
\epsilon(\phi)_{m,p}
=
\frac{M_{\rm pl}^{2}}{2f^{2}K(\phi)_{m,p}}
\,
\frac{
\left[
\sin\!\left(\frac{\phi}{f}\right)
\left(
1+\alpha\sin\!\left(\frac{\phi}{f}\right)
\right)
+
2\alpha
\left(
1+\cos\!\left(\frac{\phi}{f}\right)
\right)
\cos\!\left(\frac{\phi}{f}\right)
\right]^2
}{
\left(
1+\cos\!\left(\frac{\phi}{f}\right)
\right)^2
\left(
1+\alpha\sin\!\left(\frac{\phi}{f}\right)
\right)^2},
\end{equation}
\begin{equation}
    \mathcal{P}=1+\dfrac{3M_{\rm pl}^2\alpha^2\cos^2(\phi/f)}{2f^2\,\gamma^2}.
\end{equation}
In order to match the condition $\gamma^2>0$ we see that $ |\alpha| < 1$. In the inflationary regime and using the small-angle approximation we have $\gamma(\phi)^2\simeq 1+\alpha\frac{\phi}{f}$ and  
\begin{equation}
    K_{\rm M}\simeq\frac{1}{1+\alpha\frac{\phi}{f}}+\frac{3M_{\rm pl}^2\alpha^2}{2f^2(1+2\alpha\frac{\phi}{f})}, \quad 
    K_{\rm P}\simeq\frac{1}{1+\alpha\frac{\phi}{f}}.
\end{equation} 
Consequently, we have for the Palatini factor
\begin{equation}
    \mathcal{P}\simeq 1+\frac{3M_{\rm pl}^2\alpha^2(1+\alpha\frac{\phi}{f})}{2f^2(1+2\alpha\frac{\phi}{f})}.
    \label{eq:pala_fac_sine}
\end{equation}
For finite values of $\alpha$ and $\phi$, the correction term scales as
$f^{-2}$. Therefore $\lim_{f\to\infty}\mathcal{P}=1$.
Now, for $\alpha>0$ and $0\leq\phi/f\ll 1$, the fraction $(1+\alpha\frac{\phi}{f})/(1+2\alpha\frac{\phi}{f}) \leq 1$. And considering  $ |\alpha| < 1$ we have the bound for ${\cal P}$
\begin{equation}
\mathcal P <
1+\frac{3M_{\rm pl}^2}{2f^2}.
\end{equation}
\\
As can be seen in Fig.~\ref{fig:nsr_seno}, no significant differences 
between the metric and Palatini formalisms are observed \textbf{across the parameter space} consistent with Planck 2018 constraints.
Since the Palatini factor does not vanish for finite values of $f$, 
differences between the metric and Palatini predictions are expected 
for Coupling~B. These differences are visible in Fig.~\ref{fig:pala_sin_small}, 
where the metric (solid) and Palatini (dashed) curves are appreciably 
separated, with both predicting values of $r$ well above the Planck 2018 
and ACT DR6 confidence bands. At the smallest sampled values of $f$ (5, 6), the Palatini factor takes the maximal values 
$\mathcal{P}^5_{max}=1.06$ and $\mathcal{P}^6_{max}=1.04$.

\subsubsection{Coupling C: $\gamma^2(\phi)=1+\alpha\frac{\phi}{M_{\rm pl}}$}
\begin{figure}[H]
    \centering
    \includegraphics[width=0.6\linewidth]{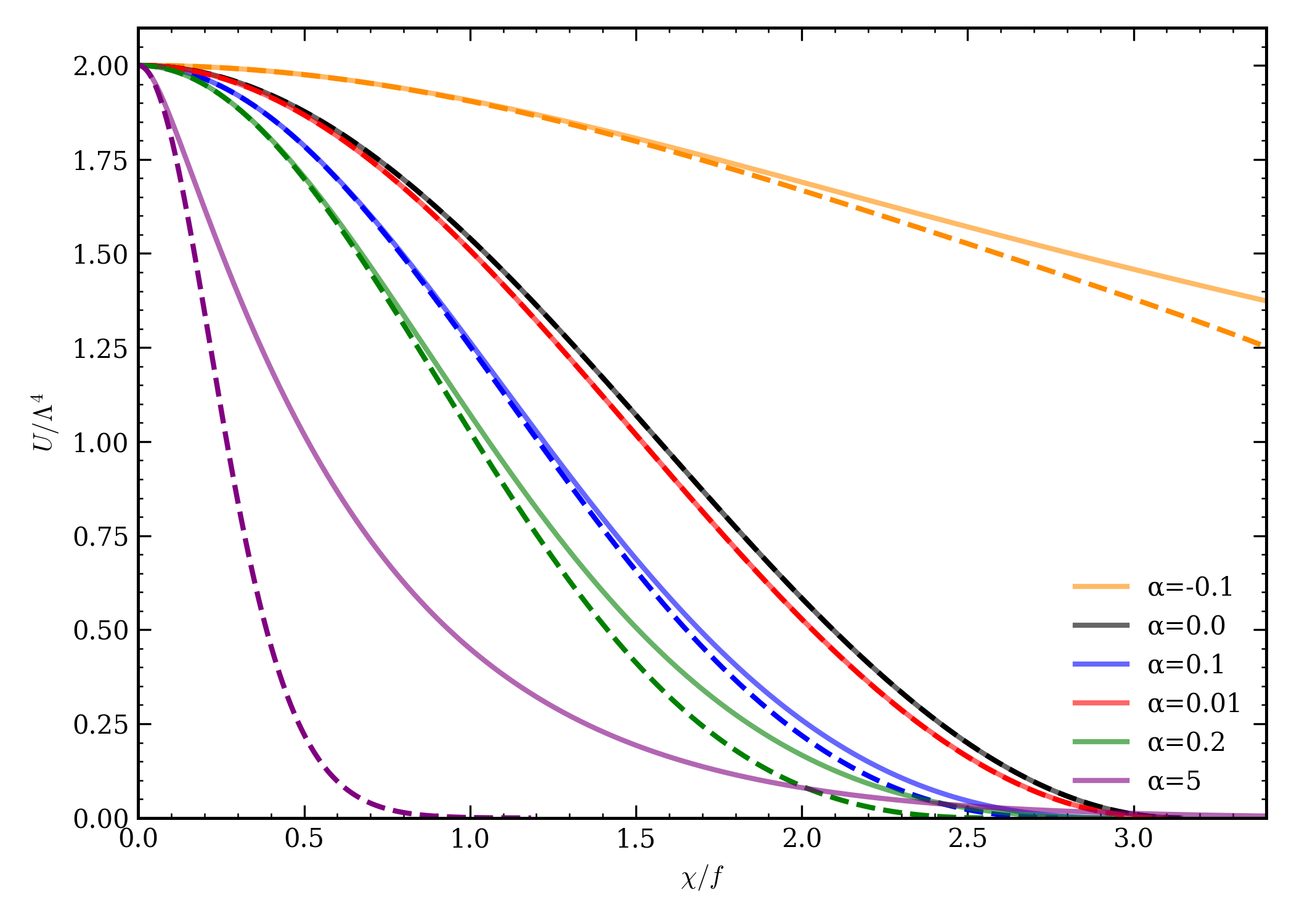}
    \caption{Potential in the Einstein frame $U/\Lambda^4$ as a function of $\chi/f$ for different values of $\alpha$ ($f=1$).}
    \label{fig:potential_comparisonC}
\end{figure}

\begin{figure}[H]
    \centering
    \begin{subfigure}[t]{0.496\textwidth}
        \centering
        \includegraphics[width=\linewidth]{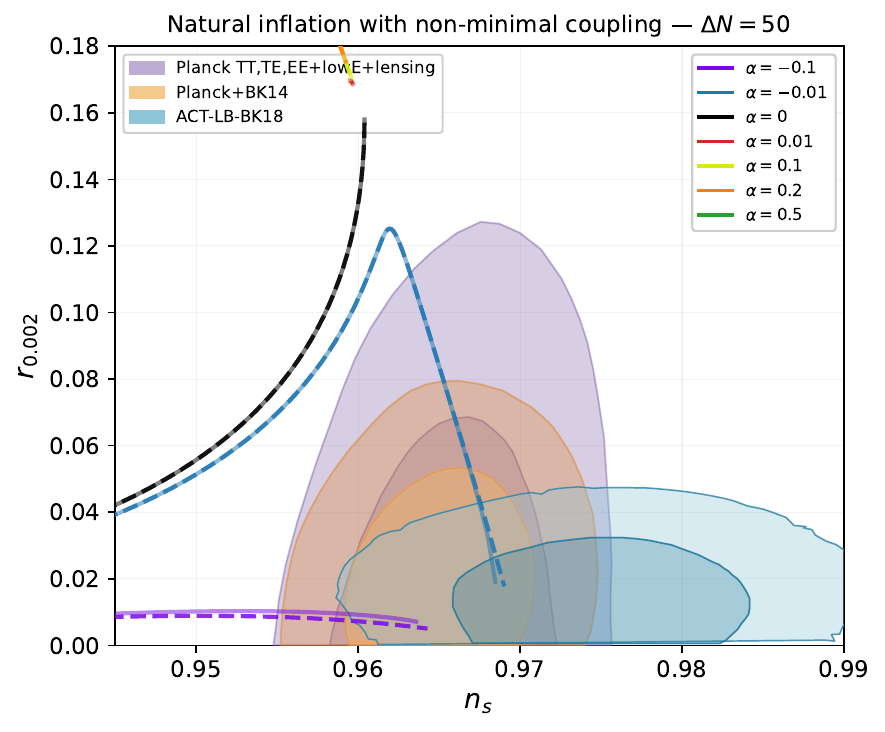}

        \label{fig:L_nsr_N50}
    \end{subfigure}
    \hfill
    \begin{subfigure}[t]{0.496\textwidth}
        \centering
        \includegraphics[width=\linewidth]{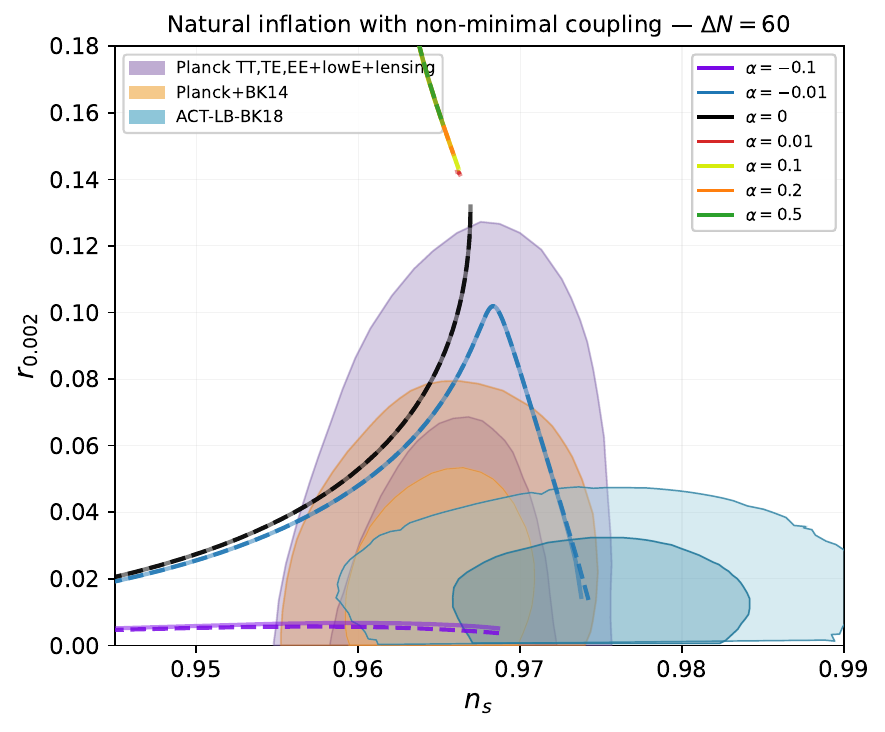}
     
        \label{fig:L_nsr_N60}
    \end{subfigure}

    
    \caption{Predictions for $n_s$ and $r$ for the linear non-minimal 
coupling  with $3 < f < 80$ for positive alphas and for negative alphas $1.7<f<\frac{1}{\pi|\alpha|}$ 
(in units of $M_{\mathrm{pl}}$). Solid and dashed lines correspond to the 
metric and Palatini formalisms. Shaded regions as in Fig.~\ref{fig:nsr_coseno}. }
    \label{fig:nsr_L}
\end{figure}

\begin{figure}[H]
    \centering
    \includegraphics[width=0.5\linewidth]{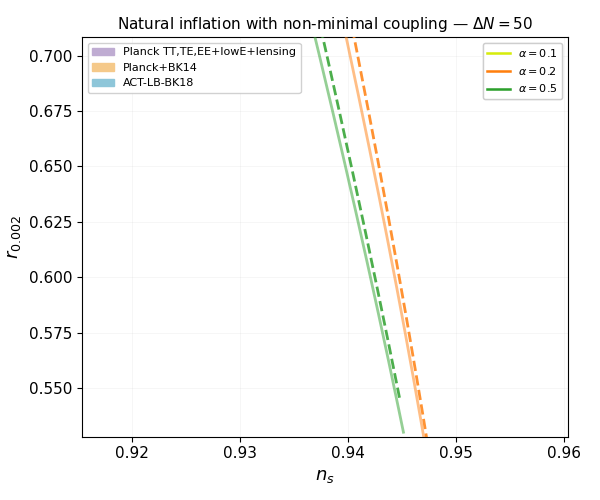}
    \caption{{Predictions for $n_s$ and $r$ for the linear non-minimal 
coupling. Solid and dashed lines correspond to the 
metric and Palatini formalisms with $2< f < 6$ 
(in units of $M_{\mathrm{pl}}$).}}

    \label{fig:pala_lin_small}
\end{figure}

This coupling explicitly violates both the shift symmetry (U(1)) and the CP symmetry of the original Natural Inflation potential. Consequently, within the framework of Natural Inflation, it should be regarded primarily as a mathematical toy model rather than a theoretically motivated extension. Nevertheless, similar linear non-minimal couplings have been shown to yield phenomenologically viable predictions in other inflationary scenarios \cite{kallosh2025actsptchaoticinflation,Kallosh:2013maa}. For this reason, it is still worthwhile to investigate their impact on the predictions of Natural Inflation and to compare the resulting metric and Palatini formulations.
\\
On the other hand, by introducing a structure that breaks both symmetries of the potential, the model becomes largely \textit{ad hoc} and loses the symmetry-based motivation that underlies natural inflation as a protected inflationary mechanism. In this case, the Einstein-frame potential takes the form, shown in Fig.~\ref{fig:potential_comparisonC}.
\noindent
and the slow-roll parameter $\epsilon$:
\begin{equation}
\epsilon(\phi)
=
\frac{M_{\rm pl}^{2}}{2\,K(\phi)_{m,p}}
\left[
\frac{\tan\!\left(\dfrac{\phi}{2f}\right)}{f}
+
\frac{2\alpha}
{M_{\rm pl}\,\gamma(\phi)^{2}}
\right]^{2}.
\end{equation}
The condition $\gamma^2>0$ implies
\begin{equation*}
    \gamma^2 = 1 + \frac{\alpha\phi}{M_{\rm pl}} > 0 \implies
    \begin{cases}
        \phi > 0 & \text{always satisfied for } \alpha > 0 \\[6pt]
        \phi < \dfrac{M_{\rm pl}}{|\alpha|} \equiv \phi_{\rm max} & \text{for } \alpha < 0.
    \end{cases}
\end{equation*}
In the inflationary regime $\gamma(\phi)^2=1+\alpha\frac{\phi}{M_{\rm pl}}$, so
\begin{equation}
    K_{\rm M}= \frac{1}{\gamma^2}+\frac{3\alpha^2}{2\gamma^4},\quad 
    K_{P}=1/\gamma^2,
\end{equation}
so
\begin{equation}
    \mathcal{P}=1+\frac{3\alpha^2}{2\gamma^2}.
\end{equation}
Since the Palatini factor achieves its maximum at $\phi \to 0$ at $\alpha>0$, we can write

\begin{equation}
    \mathcal{P}_{max} = 1 + \frac{3\alpha^2}{2},
\end{equation}
For $\alpha<0$, instead, $\gamma^2<1$ for $\phi>0$, so that $\mathcal{P}$ exceeds this value and diverges as $\phi \to M_{\rm pl}/|\alpha|$. For the values of $\alpha$ consistent with successful inflation in this model 
($|\alpha| \lesssim 0.5$), this gives

\begin{equation}
    \mathcal{P}^{\alpha=0.5}\leq 1 + \frac{3(0.5)^2}{2} = 1.375,
    \qquad
    \mathcal{P}^{\alpha=0.2}\leq 1 + \frac{3(0.2)^2}{2} = 1.06,
\end{equation}
\\
However, no significant differences between the metric and Palatini formulations are observed within the region of parameter space that yields phenomenologically relevant values of the inflationary observables, as shown in Fig.~\ref{fig:nsr_L}. This does not imply that the two formulations are identical since the Palatini factor remains different from unity. Rather, the differences are simply too small to produce an appreciable separation in the usual $(n_s,r)$ parameter space. These deviations become visible only outside the observationally relevant region, as illustrated in Fig.~\ref{fig:pala_lin_small}.

\noindent For negative $\alpha$, the small deviations observed within the shaded regions occur when the upper limit of the field grid approaches the $\gamma^2 \to 0^+$ boundary, $\phi_{\max} = \min\left(M_{\rm pl}/|\alpha|,\, \pi\right)$. Although $\gamma^2$ remains positive for the parameter values considered, its smallness leads to large values of quantities such as $K_{\rm P}$ and $J$, so the numerical evaluation becomes increasingly demanding and potentially prone to overflow as this boundary is approached. The deviations shown are therefore dominated by the physical enhancement of the kinetic sector near $\gamma^2 \to 0^+$.

\subsubsection{Coupling D: $\gamma^2(\phi)=1+\alpha\frac{\phi^2}{M_{\rm pl}^2}$}
\begin{figure}[H]
    \centering
    \includegraphics[width=0.6\linewidth]{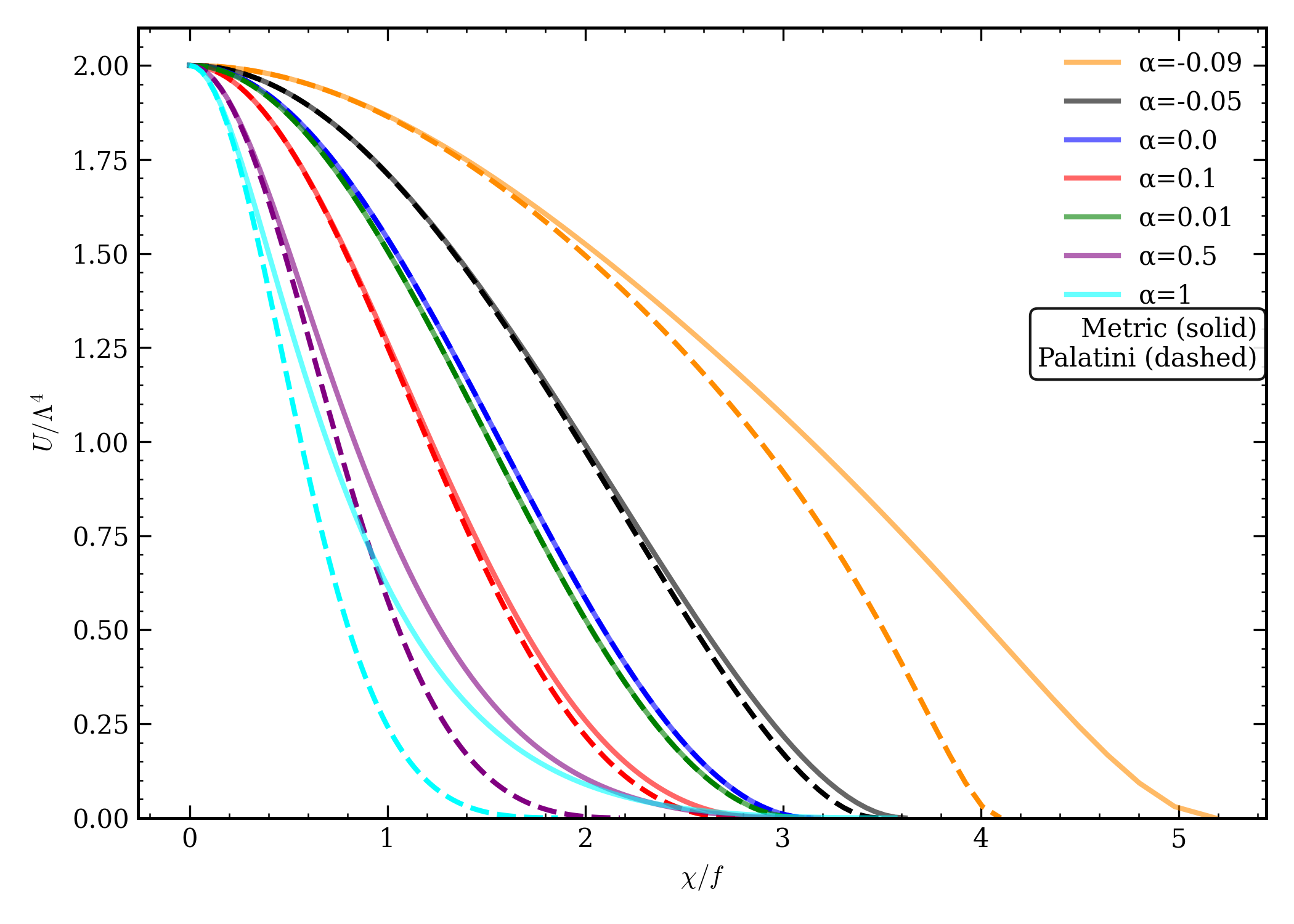}
    \caption{Potential in the Einstein frame $U/\Lambda^4$ as a function of $\chi/f$ for different values of $\alpha$ ($f=1$).}
    \label{fig:potential_comparisonD}
\end{figure}

This coupling violates the shift symmetry U(1) but preserves the CP symmetry of the original Natural Inflation potential. Despite the explicit breaking of the shift symmetry, this 
 coupling has been extensively studied in the literature in 
the context of Natural Inflation and related models 
\cite{BeltranAlmeida:2018ycz, Bostan_2023_BICEP_LINEAL, Reyimuaji_2021_LINEAL,dos_Santos_2024}. In this case, the Einstein-frame potential takes the form, shown in Fig.~\ref{fig:potential_comparisonD}.

\noindent
and the slow-roll parameter $\epsilon$:
\begin{equation}
\epsilon(\phi)
=
\frac{M_{\rm pl}^{2}}{2\,K(\phi)_{m,p}}
\left[
\frac{\tan\!\left(\dfrac{\phi}{2f}\right)}{f}
+
\frac{4\alpha\,\phi}
{M_{\rm pl}^{2}\,\gamma(\phi)^{2}}
\right]^{2}.
\end{equation}
The condition $\gamma^2>0$ is automatically satisfied for $\alpha \geq 0$, while for $\alpha<0$ it restricts the field range to $\phi < \phi_{\rm crit} \equiv M_{\rm pl}/\sqrt{|\alpha|}$. The Palatini factor is, exactly,
\begin{equation}
\mathcal{P}(\phi)=1+\frac{6\alpha^2\phi^2}{M_{\rm pl}^2\,\gamma^2}.
\end{equation}
Writing $u\equiv\alpha\phi^2/M_{\rm pl}^2$, this implies $J=6\alpha\,u/(1+u)$, so that for $\alpha>0$
\begin{equation}
J(\phi)<6\alpha
\end{equation}
for every $\phi$, independently of $f$ and $\phi_*$. For the values used here ($\alpha\lesssim10^{-2}$) this gives $J_{\max}\lesssim0.06$, and the two formalisms are nearly degenerate throughout the physically relevant domain.

\begin{figure}[H]
    \centering
    \begin{subfigure}[t]{0.496\textwidth}
        \centering
        \includegraphics[width=\linewidth]{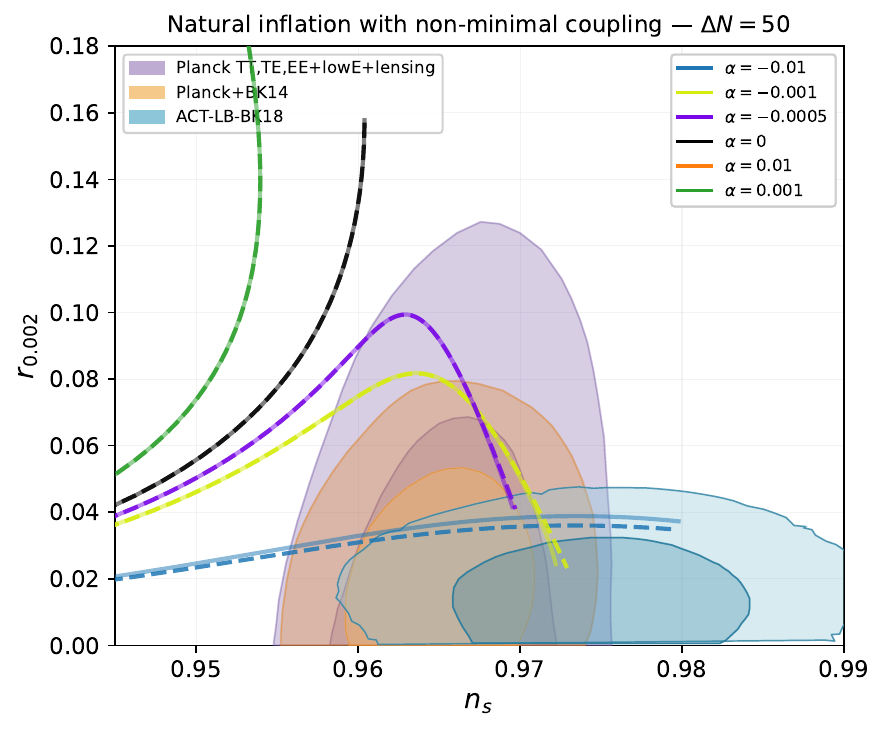}

        \label{fig:C_nsr_N50}
    \end{subfigure}
    \hfill
    \begin{subfigure}[t]{0.496\textwidth}
        \centering
        \includegraphics[width=\linewidth]{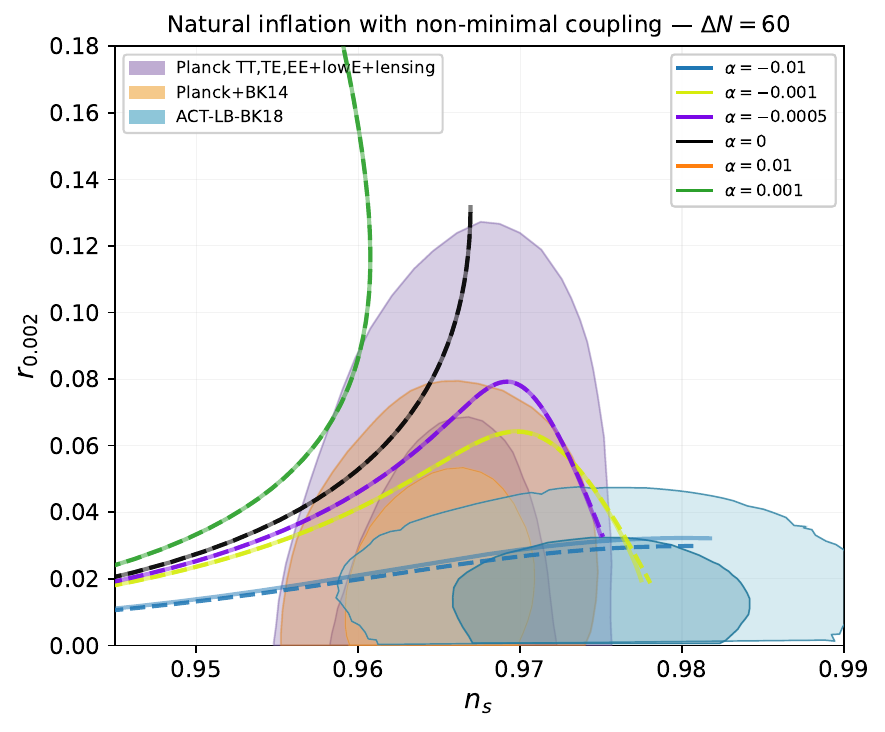}

        \label{fig:CUADRA_nsr_N60}
    \end{subfigure}
 
    \caption{Predictions for $n_s$ and $r$ for the quadratic non-minimal 
coupling. Solid and dashed lines correspond to the 
metric and Palatini formalisms  with $3 < f < 80$ for positive alphas and for negative alphas $1.7<f<\frac{1}{\pi\sqrt{|\alpha|}}$ 
(in units of $M_{\mathrm{pl}}$). Shaded regions as in Fig.~\ref{fig:nsr_coseno}. }
    \label{fig:nsr_CUADRA}
\end{figure}
For negative $\alpha$, the same behavior observed for coupling C is found near the upper boundary of the allowed field range. This behavior is associated with the proximity to the $\gamma^2\to0^+$ regime and is discussed in the corresponding analysis of coupling C.

\subsubsection{Coupling E: $\gamma^2(\phi)=1+\alpha\frac{M_{\rm pl}^2}{\phi^{2}}$}
\begin{figure}[H]
    \centering
\includegraphics[width=0.6\linewidth]{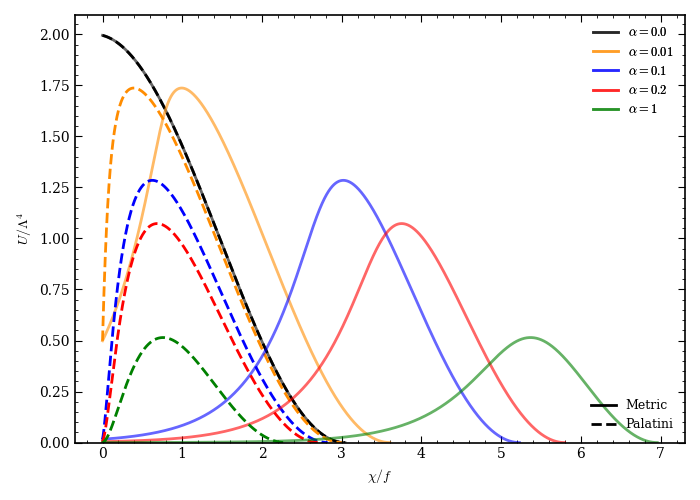}
    \caption{Potential in the Einstein frame $U/\Lambda^4$ as a function of $\chi/f$ for different values of $\alpha$ ($f=1$).\protect \footnotemark }
    \label{fig:potential_comparisonE}
\end{figure}

Finally, we analyze the coupling $\gamma^2(\phi)=1+\alpha\frac{M_{\rm pl}^2}{\phi^{2}}$.  This coupling was introduced motivated by the search for a large-field-type inflationary
behavior, in contrast with the small-field regime that characterizes the minimally coupled
model and Couplings A--D discussed above. This coupling preserves parity symmetry
but as the other power law cases explored before, it explicitly breaks the shift symmetry $U(1)$. The
coupling function develops a genuine singularity at $\phi = 0$: while $\gamma^2 > 0$ is
automatically satisfied for any $\alpha > 0$ and any $\phi \neq 0$, the point $\phi = 0$
itself must be excluded from the domain, since $\gamma^2(\phi) \to \infty$ as $\phi \to 0$. In this case, the Einstein-frame potential takes the form, shown in Fig.~\ref{fig:potential_comparisonE}.

\begin{figure}[H]
    \centering
    \begin{subfigure}[t]{0.496\textwidth}
        \centering
        \includegraphics[width=\linewidth]{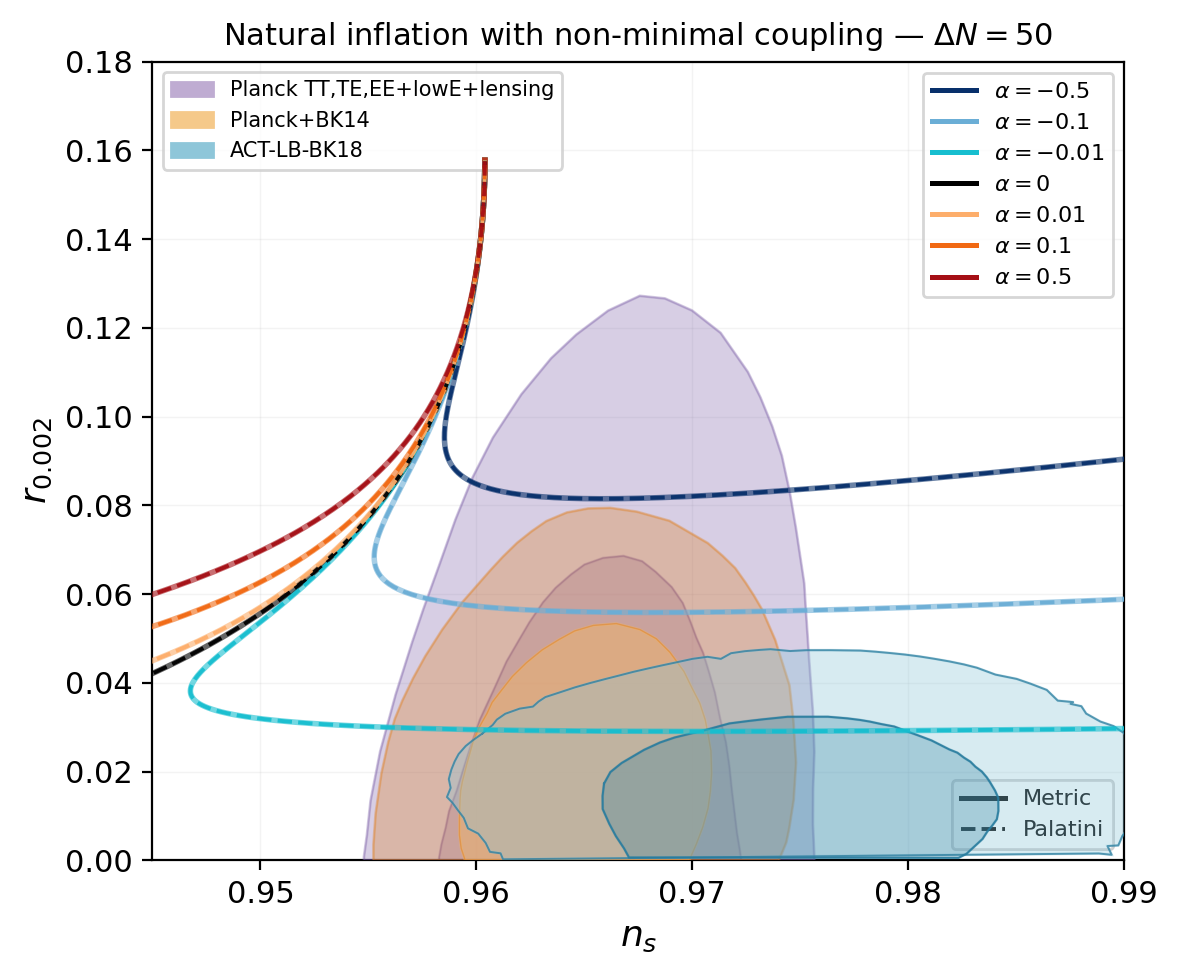}

        \label{fig:C_nsr_N50}
    \end{subfigure}
    \hfill
    \begin{subfigure}[t]{0.496\textwidth}
        \centering
        \includegraphics[width=\linewidth]{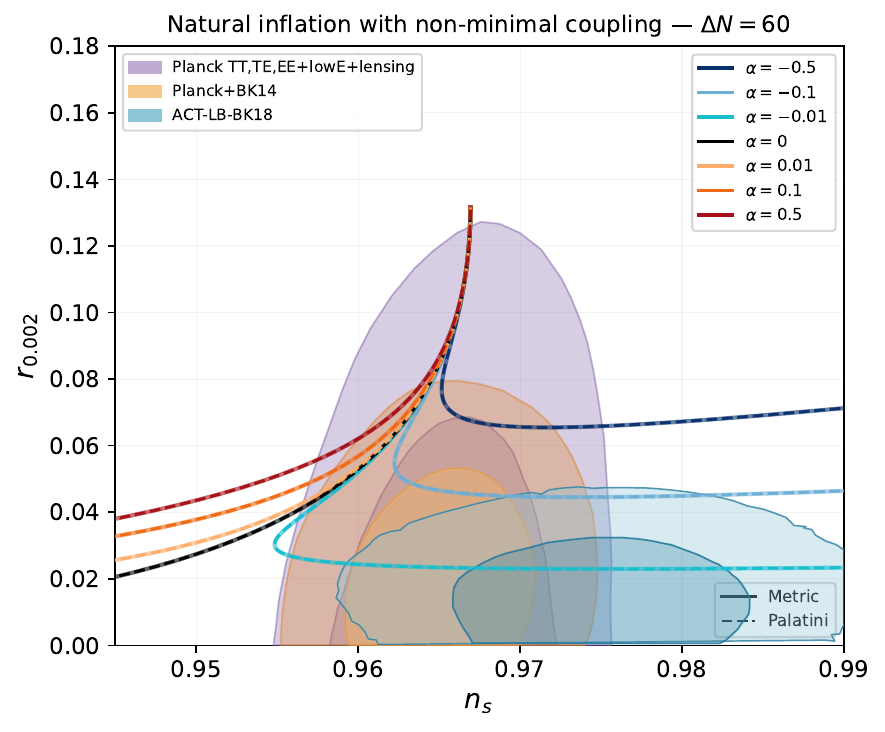}

        \label{fig:CUADRA_nsr_N60}
    \end{subfigure}

    \caption{Predictions for $n_s$ and $r$ for the inverse quadratic non-minimal
coupling $\gamma^2(\phi)=1+\alpha M_{\mathrm{pl}}^2/\phi^2$. Solid and dashed
lines correspond to the metric and Palatini formalisms with $4 < f < 80$
for both positive and negative alphas. For
$\alpha<0$ the field is restricted to $\phi>\phi_{\rm crit}=\sqrt{|\alpha|}\,M_{\mathrm{pl}}$. Shaded regions as in Fig.~\ref{fig:nsr_coseno}.}
    \label{fig:nsr_inv}
\end{figure}

\footnotetext{Unlike the other couplings where the canonical field integration starts at $\phi_{\text{init}} = 0$, here the integral defining $\chi(\phi)$ is performed
    starting from $\phi_{\text{init}} = 0.1\,M_{\rm{pl}}$, since $\chi_M(\phi)$ diverges
    logarithmically as $\phi \to 0$ for this coupling.
    This choice is a plotting convention rather than a physical starting point of inflation:
    shifting $\phi_{\text{init}}$ displaces every curve rigidly along the horizontal axis,
    without altering their relative shape by an amount
    $\Delta\chi_M = \sqrt{6}\,M_{\rm{pl}}\,\ln(\phi_{\text{init},2}/\phi_{\text{init},1})$
    between two choices of $\phi_{\text{init}}$.}
\noindent
The slow-roll parameter $\epsilon$ admits an exact closed-form expression, without further
approximation:
\begin{equation}
    \epsilon(\phi)_{m,p} = \frac{M_{\rm{pl}}^2}{2K(\phi)_{m,p}}
    \left[\frac{\tan(\phi/2f)}{f} - \frac{4\alpha M_{\rm{pl}}^2}{\phi(\phi^2 + \alpha M_{pl}^2)}\right]^2
    \label{eq:epsilon_E}
\end{equation}
and the metric and Palatini kinetic factors read
\begin{equation}
    K_{\rm M}(\phi) = \frac{1}{\gamma^2} + \frac{6\alpha^2 M_{\rm{pl}} ^6}{\phi^6 \gamma^4},
    \qquad
    K_{\rm P}(\phi) = \frac{1}{\gamma^2},\qquad
    \label{eq:K_E}
\end{equation}
\noindent
so that the Palatini factor is given exactly by
\begin{equation}
    \mathcal{P}(\phi) = \frac{K_M(\phi)}{K_P(\phi)} = 1 + \frac{6\alpha^2 M_{\rm{pl}}^6}{\phi^6 \gamma^2}
    \label{eq:P_E_exact}.
\end{equation}
Writing $u\equiv\alpha M_{\rm pl}^2/\phi^2$, the deviation from unity $J(\phi)$ is
\begin{equation}
J(\phi)=\frac{6}{\alpha}\,\frac{u^3}{1+u}.
\end{equation}
Unlike coupling D, $J$ has no global bound, since it diverges as $\phi\to0$. However, slow-roll inflation is excluded near the origin  and requires $\phi_*$ large, where $u_*\ll1$ and
\begin{equation}
J_*\simeq6\alpha^2\left(\frac{M_{\rm pl}}{\phi_*}\right)^6.
\label{J*_E}
\end{equation}
Since the field rolls towards larger $\phi$, $u=\alpha M_{\rm pl}^2/\phi^2$ decreases along the trajectory and is maximal at $\phi_*$. For $u_*\ll1$, i.e.\ $\phi_*^2\gg\alpha M_{\rm pl}^2$, one has $\gamma^2\simeq1$, $U\simeq V$ and $K_{\rm M}\simeq K_{\rm P}\simeq1$ over the entire inflationary trajectory, together with $J\leq J_*\simeq6\alpha^2(M_{\rm pl}/\phi_*)^6\ll1$. In this regime no differences between the two formalisms are expected, and both reduce to minimally coupled Natural Inflation. Interestingly enough, particular choices of negative $\alpha$ make the model compatible with current observational bounds as shown in Fig. ~\ref{fig:nsr_inv}.

\subsubsection{Summary of results 
} We conclude this section summarizing our findings. Table~\ref{tab:J_summary} collects the exact $J=\mathcal{P}-1$ for each coupling, together with its bound on the physical domain and the parameter that suppresses it.

\begin{table}[H]
\centering
\small
\renewcommand{\arraystretch}{1.6}
\begin{tabular}{l l l l l}
\toprule
Coupling & $\gamma^2(\phi)$ & $J(\phi)$ & Bound ($\alpha>0$) & Suppressed by \\
\midrule
A & $1+\alpha\big(1+\cos\frac{\phi}{f}\big)$ & $\dfrac{3M_{\rm pl}^2\alpha^2\sin^2(\phi/f)}{2f^2\gamma^2}$ & $J\leq\dfrac{3\alpha^2M_{\rm pl}^2}{2f^2}$ & $f$ \\
B & $1+\alpha\sin\frac{\phi}{f}$ & $\dfrac{3M_{\rm pl}^2\alpha^2\cos^2(\phi/f)}{2f^2\gamma^2}$ & $J\leq\dfrac{3\alpha^2M_{\rm pl}^2}{2f^2(1-|\alpha|)}$ & $f$ \\
C & $1+\alpha\dfrac{\phi}{M_{\rm pl}}$ & $\dfrac{3\alpha^2}{2\gamma^2}$ & $J\leq\dfrac{3\alpha^2}{2}$ & $\alpha$ \\
D & $1+\alpha\dfrac{\phi^2}{M_{\rm pl}^2}$ & $\dfrac{6\alpha^2\phi^2}{M_{\rm pl}^2\gamma^2}$ & $J<6\alpha$ & $\alpha$ \\
E & $1+\alpha\dfrac{M_{\rm pl}^2}{\phi^2}$ & $\dfrac{6\alpha^2M_{\rm pl}^6}{\phi^6\gamma^2}$ & $J_*\simeq6\alpha^2\Big(\dfrac{M_{\rm pl}}{\phi_*}\Big)^6$ & $\phi_*$ \\
\bottomrule
\end{tabular}
\caption{Exact $J=\mathcal{P}-1$ for each coupling, its bound for $\alpha>0$, and the parameter controlling its suppression. For B the bound holds for all $|\alpha|<1$. For $\alpha<0$ in C and D, $\gamma^2\to0$ at the edge of the physical domain and $J$ is unbounded there; for E with $\alpha<0$ the domain is $\phi>\sqrt{|\alpha|}\,M_{\rm pl}$.}
\label{tab:J_summary}
\end{table}

\section{Relation to $\alpha$-attractors}
\label{sec:alpha_attractors}

In this section we comment on the connection  between $\alpha$-attractors 
and $\xi$-attractors pointed out in 
ref.~\cite{Galante_2015,J_rv_2018}\footnote{Throughout this section, 
the symbol $\xi$ denotes the non-minimal coupling 
parameter, which corresponds to $\alpha$ in the 
notation used in the rest of this work. This 
distinction is made to avoid confusion with the 
$\alpha$-attractor parameter, denoted $\alpha_{\rm attr}$ 
hereafter.}. Let us start with a Jordan frame action of the form
\begin{equation}
S_J=\int d^4x\sqrt{-\tilde{g}}
\left[
\frac{M_{\rm {pl}}^2}{2}\gamma(\phi)^2\tilde{R} 
-\frac{1}{2}h(\phi)\tilde{g}^{\mu\nu}\partial_\mu\phi\partial_\nu\phi
-V(\phi)
\right],
\label{salphaatt}
\end{equation}
in which a kinetic function $h(\phi)$ is introduced from the start. Going to Einstein frame we get
\begin{equation}
S_E=\int d^4x\sqrt{- g}
\left[
\frac{M_{\rm {pl}}^2}{2} R
-\frac{1}{2}K(\phi) g^{\mu\nu}\partial_\mu\phi\partial_\nu\phi
-U(\phi)
\right].
\end{equation}
where now, the kinetic factor includes the function $h$  
\begin{equation}
K(\phi)=\frac{h(\phi)+6M_{\rm pl}^2{\rm p}\gamma'^2}{\gamma^2},
\end{equation}
where ${\rm p} = 1, 0$ for the metric and Palatini formulation  respectively. Choosing $h(\phi) = M_{\rm {pl}}^2  \gamma'^2/\xi$ with $\xi$ the (dimensionless) non-minimal coupling parameter, we get
\begin{equation}
K(\phi)=6 M_{\rm {pl}}^2 \left(\frac{1}{6\xi} +  {\rm p}\right)  \frac{\gamma'^2}{\gamma^2} = 6 M_{\rm {pl}}^2 \left(\frac{1}{6\xi} +  {\rm p}\right)  \left( \frac{\partial\ln \gamma}{\partial\phi}\right)^2 ,
\end{equation}
where we see that the kinetic term acquires a pole at $\gamma^2=0$. The structure of the pole $1/\gamma^2$ depends on the particular form of the coupling function $\gamma$.  Here, following \cite{Galante_2015} we  identify 
\begin{equation}\label{alphaxi}
\alpha_{\rm attr} \equiv  \left(\frac{1}{6\xi} +  {\rm p}\right)  ,
\end{equation}
so, we can write the Einstein frame action as 
\begin{equation} \label{actionalpha}
S_E=\int d^4x\sqrt{- g}
\left[
\frac{M_{\rm {pl}}^2}{2} R
-\frac{3 \alpha_{\rm attr}}{4} \left(\frac{\partial_{\phi} \gamma^2}{\gamma^2}\right)^2 g^{\mu\nu}\partial_\mu\phi\partial_\nu\phi
-U(\phi)
\right],
\end{equation}
hence, the kinetic coefficient $\alpha_{\rm attr}$ depends on 
both the non-minimal coupling parameter $\xi$ and 
the gravitational formulation through ${\rm p}$. 
The canonical field can be obtained from
\begin{equation}
\frac{d\chi}{d\phi}=\sqrt{K(\phi)} = \sqrt{6 \alpha_{\rm attr}}M_{\rm pl}\frac{\partial\ln \gamma(\phi)}{\partial\phi},
\label{canonicalalpha}
\end{equation}
which can be easily integrated to
\begin{equation}
 \chi  = \sqrt{6 \alpha_{\rm attr}}M_{\rm pl} \,  \ln \gamma (\phi).
 \end{equation}
The inverse is also easily obtained in terms of the inverse of the $\gamma$ function:
\begin{equation}
\gamma(\phi) =e^{\frac{\chi}{\sqrt{6 \alpha_{\rm attr}}M_{\rm pl}}}\; \xrightarrow{}\;\phi=\gamma^{-1}\left( e^{\frac{\chi}{\sqrt{6 \alpha_{\rm attr}} M_{\rm pl}}} \right).
\end{equation} 
\subsection{Pole structure and the order/residue split}

To make the attractor mechanism explicit, introduce the variable
\begin{equation}
\rho \equiv \frac{1}{\gamma^2(\phi)}, \qquad
\frac{\gamma'}{\gamma} = -\frac{1}{2}\frac{\rho'}{\rho},
\end{equation}
so that $(\partial_\phi\ln\gamma)^2 (\partial\phi)^2 = \tfrac{1}{4}(\rho'/\rho)^2(\partial\phi)^2$. Using $(\partial\phi)^2 = (\partial\rho)^2/(\rho')^2$, the kinetic term in Eq.~\eqref{actionalpha} becomes, in terms of $\rho$,
\begin{equation}
K_{\rm E}(\rho) = \frac{3\alpha_{\rm attr}}{2}\,\frac{M_{\rm pl}^2}{\rho^2},
\label{keroof}
\end{equation}
with no subleading terms: a single pole of order $s=2$ located at $\rho=0$ (i.e. $\gamma^2\to\infty$), with residue $a_2 = \tfrac{3}{2}\alpha_{\rm attr}M_{\rm pl}^2$. This is exactly the ``special attractor'' construction of \cite{Galante_2015}, and the identification $a_2\propto\alpha_{\rm attr}$ is where the formalism enters: \emph{the order of the pole ($s=2$) is the same in both formalisms, but the residue $a_2$ is not}, since $\alpha_{\rm attr}=1/6\xi+{\rm p}$ \cite{J_rv_2018}.

For a potential smooth and non-vanishing at the pole, $U(\rho)=U_0(1+c\rho+\dots)$ with $U_0\neq0$, the general pole argument of \cite{Galante_2015} gives, for $s=2$,
\begin{equation}
n_s = 1-\frac{2}{N}, \qquad r = \frac{8a_2}{N^2} = \frac{12\,\alpha_{\rm attr}}{N^2}.
\label{nsrpole}
\end{equation}
The spectral index depends only on the order of the pole and is therefore \emph{formalism-independent}; the tensor-to-scalar ratio depends on the residue through $\alpha_{\rm attr}$ and is therefore \emph{formalism-dependent}. In the strong-coupling limit $\xi\to\infty$,
\begin{equation}
\alpha_{\rm attr}^{M}\to 1 \;\Rightarrow\; r\to\frac{12}{N^2}\ \text{(Starobinsky point)}, \qquad
\alpha_{\rm attr}^{P}\to 0 \;\Rightarrow\; r\to 0,
\end{equation}
for both formulations. Equivalently, the Palatini factor for this construction is
\begin{equation}
\mathcal{P} \equiv \frac{K_{\rm M}}{K_{\rm P}} = \frac{1/6\xi+1}{1/6\xi} = 1+6\xi,
\end{equation}
which diverges precisely in the strong-coupling regime where the metric formalism saturates its attractor value. We stress that this is a stronger and more precise statement than ``the pole structure is the same in both formalisms'': the two formalisms share the \emph{universality class} of $n_s$, but not the attractor prediction for $r$. Now, specifically, for natural inflation with a periodic coupling function $\gamma$, the kinetic term doesn't develop a pole structure in the weak-field limit that would flatten the potential near the pole and yield robust inflationary predictions independent of the potential's details. Consequently, this approach is not useful for bringing natural inflation to agreement with observations. In order to improve the model, one would need to add extra degrees of freedom, for instance additional inflationary scalar fields with a large-field potential. Such modifications have been explored in the literature, for example by combining a potential of the form $\lambda\phi^4$ with a pole of the form $\gamma^{-2}\sim(1-\phi^2)^{-2}$. This kind of modification produces a successful model of inflation with suppressed tensor perturbations, without generating observable non-Gaussianities or isocurvature perturbations at CMB scales~\cite{McDonough:2020gmn}. A number of works have studied the inflationary dynamics and observable predictions of such models; see for instance \cite{McDonough:2020gmn, Linde:2018hmx, Lorenzoni:2024krn}. Based on the conclusions reached here, the two formalisms do not imply any difference in the pole structure of the models, only a different relation between the $\alpha$ and $\xi$ parameters in each formalism. Palatini $\xi$-attractor constructions remain an active area of study; see e.g.~\cite{Dioguardi:2026} for a recent, structurally distinct realization based on a direct coupling $\gamma^2=1+\xi f(\phi)$ shared with the potential $V=V_0f(\phi)^2$, without the pole structure discussed here. We expect to return to these issues and extend our analysis to multifields models of inflation elsewhere. 
\\




\section{General Results }
\label{sec:general_results}
Based on the particular cases studied in section \ref{sec:Natural_periodic}, in this section we extend the analysis to general coupling functions for small, single field models of inflation.

\subsection{Some General Instances}
We now illustrate the classification above with three representative families of coupling functions, {periodic $\cos$ and $\sin$ and power law.}  

\paragraph{Cosine-type couplings, $g=g_1(\cos(\phi/f))$.} Differentiating,
\begin{equation}
g_{,\phi} = -\frac{\sin(\phi/f)}{f}\,g_1'(\cos(\phi/f)),
\end{equation}
so that $g_{,\phi}(0;f)=0$ identically, for any smooth $g_1$: this family always satisfies the case I condition, with the field-driven suppression $J_*\propto\phi_*^2/f^4$ arising purely from the vanishing of $\sin(\phi_*/f)$ as $\phi_*/f\to0$, independently of the specific form of $g_1$. A caveat applies: this suppression assumes the coefficient $[g_1'(1)]^2/g_1(1)$ remains bounded as the coupling's internal parameters (e.g.\ $\alpha$) vary, and may fail if it grows without bound.

\paragraph{Sine-type couplings, $g=g_2(\sin(\phi/f))$.} Differentiating $g$ we have
\begin{equation}
g_{,\phi} = \frac{\cos(\phi/f)}{f}\,g_2'(\sin(\phi/f)),
\end{equation}
so that $g_{,\phi}(0;f)=g_2'(0)/f$, generically non-zero, giving
\begin{equation}
J_* \approx \frac{3M_{\rm pl}^2\,[g_2'(0)]^2}{2f^2\,g_2(0)}
\end{equation}
Since this value is independent of $\phi_*$, the field-driven suppression of case I is absent: $J_*$ does not shrink as $\phi_*\to0$. Whether the coupling falls into Case I or Case II instead depends on the size of $g_{2,y}(0)^2/g_2(0)$ relative to $f^2$: if this ratio is well-behaved, large $f$ places the coupling in case I through a parameter-driven suppression, rather than the field-driven mechanism above.

\paragraph{Power-law couplings, $g=1+\alpha(\phi/M_{\rm pl})^n$.} Differentiating $g$ we have
\begin{equation}
g_{,\phi} = \frac{n\alpha}{M_{\rm pl}}\left(\frac{\phi}{M_{\rm pl}}\right)^{n-1}.
\end{equation}
The cases $n=0$ (constant coupling, $J\equiv0$ identically) and $n=1$ (linear coupling, already treated separately as Coupling C) are trivial and excluded from the discussion below.\\
\textbf{Case $n\geq2$:} $g_{,\phi}(0;\alpha)=0$ for any $\alpha$, since $n-1\geq1$, so this family falls into case I, with $J_*\propto\phi_*^{2(n-1)}$ and no additional condition on $f$, unlike the cosine case above.\\
\textbf{Case $n\leq-1$:} the coupling diverges at $\phi=0$, so the Taylor classification above does not apply. Near the origin $\gamma^2\simeq\alpha(M_{\rm pl}/\phi)^{|n|}$. The exact $J$ is
\begin{equation}
J=\frac{3n^2\alpha^2}{2}\,\frac{(M_{\rm pl}/\phi)^{2|n|+2}}{1+\alpha(M_{\rm pl}/\phi)^{|n|}}\;\simeq\;\frac{3n^2\alpha^2}{2}\left(\frac{M_{\rm pl}}{\phi_*}\right)^{2|n|+2}
\end{equation}
for $\phi_*^{|n|}\gg\alpha M_{\rm pl}^{|n|}$. In that regime $\gamma^2\simeq1$, $J_*\ll1$, and both formalisms are like Natural Inflation. Coupling E ($n=-2$) gives $J_*\simeq6\alpha^2(M_{\rm pl}/\phi_*)^6$, in agreement with Eq.~\eqref{J*_E}.

\noindent For both cases, when  $\alpha<0$, $\gamma^2$ can vanish and $J\to\infty$, therefore the suppression is no longer guaranteed.

\section{Conclusions and Discussion}
\label{sec:conclusions}

We have studied Natural Inflation with non-minimal couplings to gravity in the metric and Palatini formalisms. To compare them we used the Palatini factor $\mathcal{P}=K_{\rm M}/K_{\rm P}$ and its deviation from unity $J=\mathcal{P}-1$, which measures the relative enhancement of the kinetic sector in the metric formulation. At fixed field value $\epsilon_{\rm P}=\mathcal{P}\,\epsilon_{\rm M}$, so that $J=0$ implies exact degeneracy. Since the horizon-crossing values $\phi_*^{(\rm M)}$ and $\phi_*^{(\rm P)}$ generally differ, $J$ is a diagnostic of whether differences can arise rather than a direct measure of their size in the observables.

We analyzed five coupling functions ---cosine (A), sine (B), linear (C), $\phi^2$ (D) and $\phi^{-2}$ (E)--- against Planck 2018~\cite{Akrami2020} and ACT DR6~\cite{ACT1,ACT2}. At leading order, the metric--Palatini difference is controlled by the first derivative of $g=\gamma^2$ at horizon crossing, which leads to a model-independent classification in the small-field regime $\phi_*/f\ll1$:

\textbf{Case~I}, $J_*\ll1$: both formalisms are nearly degenerate. The suppression can be field-driven, when $g_{,\phi}(0)=0$ and $J_*\propto\phi_*^2$, as for cosine-type couplings and power laws with $n\geq2$ (Coupling~A follows from $\phi_*/f\ll1$; for Coupling~D, $J<6\alpha$ is set by small $\alpha$); or parameter-driven, when $g_{,\phi}(0)\neq0$ but the residual $J_*\propto g_{,\phi}(0)^2/g(0)$ is made small by the model parameters, as for sine-type couplings at large $f$ (Coupling~B) and the linear coupling at small $\alpha$ (Coupling~C).

\textbf{Case~II}, $J_*$ Non-vanishin J: residual metric--Palatini differences survive at horizon crossing. This requires $g_{,\phi}(0)\neq0$ without parameter suppression, as for Coupling~B at small $f$, which however lies outside the observationally favored region.

Consequently, in all cases studied, the regions compatible with the data have small $J_*$, and switching the gravitational formulation does not improve the observational viability of Natural Inflation. Metric--Palatini differences do appear, but only outside the observationally favored region (Figs.~\ref{fig:pala_sin_small} and \ref{fig:pala_lin_small}, and the outer branches of Figs.~\ref{fig:nsr_seno}, \ref{fig:nsr_L} and \ref{fig:nsr_CUADRA}).

Agreement with Planck 2018 and ACT DR6 is instead achieved by negative $\alpha$ in the power-law couplings C, D and E, independently of the formalism. This comes at a price: these couplings break the periodic shift symmetry of the Nambu--Goldstone construction (Coupling~C also breaks CP), so the resulting models lose the symmetry-based motivation of Natural Inflation and should be regarded as \textit{ad hoc} extensions. Moreover, for C and D the favored region lies close to the $\gamma^2\to0^+$ boundary, where the effective Planck mass becomes small.

Natural extensions of this work include non-minimally coupled multi-field models, where curved trajectories and turns in field space may be sensitive to the gravitational formulation, and the post-inflationary dynamics, including reheating and particle production. We plan to address these topics in future work.

\section*{Acknowledgments}
JPBA acknowledges partial support from the Faculty of Science  at Universidad Nacional de Colombia under project Hermes 65484.  


\bibliography{references}  

\newpage

\appendix

\section{Numerical exploration}
\label{app:numerical}

In this appendix we describe the numerical procedure used to obtain
the inflationary observables $(n_s,r)$ reported in
Section~\ref{sec:Natural_periodic} for both the metric and Palatini formalisms,
together with the consistency checks performed to validate the
implementation.  The code is written in \textsc{Python} using
\textsc{NumPy}, \textsc{SciPy} and \textsc{Joblib}

\subsection{Pipeline}
\label{app:pipeline}

For each coupling function $\gamma^2(\phi)$, each parameter pair
$(\alpha,f)$, each target number of $e$-folds
$N_*\in\{50,60,65\}$, and each formalism (metric or Palatini), the
procedure is as follows.

\begin{enumerate}

\item \textbf{Background functions.}
  We evaluate the Einstein-frame potential
  $U(\phi)=V(\phi)/\gamma^4(\phi)$ and the kinetic factor
  $K(\phi)$.  For each coupling, $K(\phi)$ and
  $\epsilon(\phi)$ are obtained from closed-form expressions
  derived analytically; no numerical differentiation of
  $\gamma(\phi)$ is performed.  The explicit expressions are
  collected in Table~\ref{tab:couplings}, and the only quantity
  that changes between formalisms is $K(\phi)$:
  \begin{equation}
  K_{\rm M}(\phi)
    =\frac{1}{\gamma^2}
    +\frac{6M_{\rm Pl}^2\,\gamma'^2}{\gamma^2},
  \qquad
  K_{\rm P}(\phi)=\frac{1}{\gamma^2}.
  \end{equation}

\item \textbf{Slow-roll parameters.}
  The first slow-roll parameter $\epsilon(\phi)$ is evaluated from
  the coupling-specific analytic formula
  (Table~\ref{tab:couplings}).  The second parameter $\eta(\phi)$
  is obtained via centred finite differences of $U(\phi)$ and
  $K(\phi)$ on the grid:
  \begin{equation}
  \eta(\phi)=\frac{M_{\rm Pl}^2}{U}\!\left(
    \frac{U_{,\phi\phi}}{K}
    -\frac{K_{,\phi}\,U_{,\phi}}{2K^2}\right),
  \end{equation}
  where $U_{,\phi\phi}$ and $K_{,\phi}$ are computed with
  second-order central differences on the $\phi$ grid.

\item \textbf{End of inflation.}
  We locate $\phi_{\rm end}$ as the smallest root of
  $\epsilon(\phi)=1$ on the inflationary branch, using Brent's
  method (\texttt{scipy.optimize.root\_scalar} with
  \texttt{method='brentq'}).  The bracket is restricted to the
  first crossing from below on the monotonic slow-roll region,
  excluding spurious roots near the potential maximum.

\item \textbf{Horizon crossing.}
  The number of $e$-folds from $\phi$ to $\phi_{\rm end}$ is
  \begin{equation}
  N(\phi)=\frac{1}{M_{\rm Pl}}
    \int_{\phi}^{\phi_{\rm end}}
    \sqrt{\frac{K(\phi')}{2\,\epsilon(\phi')}}\;d\phi',
  \label{eq:Nint}
  \end{equation}
  evaluated with adaptive Gauss--Kronrod quadrature
  (\texttt{scipy.integrate.quad}).  The pivot value $\phi_*$ is
  found by solving $N(\phi_*)=N_*$ with Brent's method.

\item \textbf{Observables.}
  Finally,
  \begin{equation}
  r=16\,\epsilon(\phi_*),\qquad
  n_s=1-6\,\epsilon(\phi_*)+2\,\eta(\phi_*),
  \end{equation}
  where $\epsilon_*$ and $\eta_*$ at $\phi_*$ are obtained by
  linear interpolation on the precomputed grid.  Since $\phi_*$ is
  determined separately for each formalism (because $K$ enters
  Eq.~\eqref{eq:Nint}), the metric and Palatini predictions are
  compared at their respective horizon-crossing points.

\end{enumerate}

The outer loop scans the parameter space $(\alpha,f)$ for each
coupling; for each $\alpha$ the decay-constant range is adapted to
avoid the unphysical region $\gamma^2\le 0$ (see
Section~\ref{app:domain}).  The independent evaluations are
distributed across CPU cores using \textsc{Joblib}.

\subsection{Pseudocode}
\label{app:pseudocode}

Algorithm~\ref{alg:pipeline} summarises the computation of
$(n_s,r)$ for a single parameter point.
Algorithm~\ref{alg:scan} describes the outer parameter-space scan.
The coupling-dependent ingredients --- $\gamma^2(\phi)$, $K(\phi)$,
and $\epsilon(\phi)$ --- are collected in
Table~\ref{tab:couplings}; they are the \emph{only} quantities that
change between models.

\begin{algorithm}[H]
\caption{Computation of $(n_s,r)$ for a single
  $(\alpha,f,N_*,\text{formalism})$.}
\label{alg:pipeline}
\begin{algorithmic}[1]

\Require $\alpha,\;f,\;N_*,\;
  \text{formalism}\in\{\text{M},\text{P}\}$

\Statex
\State \textbf{Set up grid:}
\If{$\alpha<0$}
  \State $\phi_{\max}\gets
    \min\!\bigl(\phi_{\max}^{\gamma^2>0},\;f\pi\bigr)-\delta$
    \Comment{avoid $\gamma^2\le0$}
\Else
  \State $\phi_{\max}\gets f\pi-\delta$
\EndIf
\State $\boldsymbol{\phi}\gets
  \textsc{Linspace}\bigl(\delta,\;\phi_{\max},\;
  N_{\rm grid}\bigr)$
  \Comment{$N_{\rm grid}=10\,000$;\; $\delta=10^{-2}$}

\Statex
\State \textbf{Evaluate background on grid:}
\State $U_i \gets V(\phi_i)/\gamma(\phi_i)^4$
  \quad for all $i$
\State $K_i \gets K(\phi_i)$
  \Comment{closed-form; M or P}
\State $\epsilon_i \gets
  \epsilon_{\rm analytic}(\phi_i)$
  \Comment{closed-form; Table~\ref{tab:couplings}}
\State $\eta_i \gets$ centred finite differences
  of $U$, $K$ on grid

\Statex
\State \textbf{End of inflation:}
\State Find smallest $i$ such that
  $\epsilon_i\ge1$;
  set bracket $[\phi_{i-1},\phi_i]$
\State $\phi_{\rm end}\gets
  \textsc{Brentq}\bigl(\epsilon(\phi)-1,\;
  [\phi_{i-1},\phi_i]\bigr)$

\Statex
\State \textbf{Total $e$-folds check:}
\State $N_{\max}\gets\dfrac{1}{M_{\rm Pl}}
  \displaystyle\int_{\phi_0}^{\phi_{\rm end}}
  \sqrt{\dfrac{K}{2\epsilon}}\;d\phi$
  \quad\Comment{\textsc{Quad}}
\If{$N_{\max}<N_*$}
  \State \Return \texttt{NaN}
  \Comment{insufficient inflation}
\EndIf

\Statex
\State \textbf{Horizon crossing:}
\State Define
  $\;g(\phi)\equiv N(\phi,\phi_{\rm end})-N_*$
\State $\phi_*\gets
  \textsc{Brentq}\bigl(g,\;
  [\phi_0,\phi_{\rm end}]\bigr)$
  \Comment{tolerances: $10^{-12}$}

\Statex
\State \textbf{Observables:}
\State $\epsilon_*\gets
  \textsc{Interp}
  (\phi_*,\boldsymbol{\phi},\boldsymbol{\epsilon})$
\State $\eta_*\gets
  \textsc{Interp}
  (\phi_*,\boldsymbol{\phi},\boldsymbol{\eta})$
\State $r\gets 16\,\epsilon_*$
\State $n_s\gets 1-6\,\epsilon_*+2\,\eta_*$
\State \Return $(n_s,\,r)$

\end{algorithmic}
\end{algorithm}

\begin{algorithm}[H]
\caption{Parameter-space scan (outer loop).}
\label{alg:scan}
\begin{algorithmic}[1]

\Require List of $\alpha$ values; for each $\alpha$,
  range $[f_{\min},f_{\max}]$;
  targets $N_*\in\{50,60,65\}$

\For{each $\alpha$}
  \State $\mathbf{f}\gets
    \textsc{Linspace}(f_{\min},f_{\max},N_f)$
    \Comment{$N_f=1000$}
  \For{each $f\in\mathbf{f}$
    \textbf{in parallel}}
    \For{formalism $\in
      \{$metric, Palatini$\}$}
      \For{$N_*\in\{50,60,65\}$}
        \State $(n_s,r)\gets$
          \textsc{Algorithm~\ref{alg:pipeline}}%
          $(\alpha,f,N_*,\text{formalism})$
        \State Write
          $(\alpha,f,\text{formalism},N_*,n_s,r)$
          to CSV
      \EndFor
    \EndFor
  \EndFor
\EndFor

\end{algorithmic}
\end{algorithm}

\subsection{Coupling-dependent ingredients}
\label{app:ingredients}

Table~\ref{tab:couplings} collects the closed-form expressions for
 $\gamma^2(\phi)$. $K_M(\phi)$, 
and $\epsilon(\phi)$ are
hardcoded for each coupling.  Every other step of the pipeline
(root-finding, quadrature, interpolation, output) is identical
across models.  In all cases $K_P=1/\gamma^2$.

\begin{table}[h]
\centering
\renewcommand{\arraystretch}{2.0}
\small
\begin{tabular}{@{}clc@{}}
\toprule
 & $\gamma^2(\phi)$ & Domain \\
\midrule
A & $1+\alpha\bigl(1+\cos\tfrac{\phi}{f}\bigr)$
  & $\alpha>-\tfrac{1}{2}$ \\
B & $1+\alpha\sin\tfrac{\phi}{f}$
  & $|\alpha|<1$ \\
C & $1+\alpha\,\dfrac{\phi}{M_{\rm Pl}}$
  & $\gamma^2>0$ \\
D & $1+\alpha\,\dfrac{\phi^2}{M_{\rm Pl}^2}$
  & $\gamma^2>0$ \\
\bottomrule
\end{tabular}
\caption{Coupling functions and their validity domains.
  }
\label{tab:couplings}
\end{table}

The metric kinetic factors, obtained by substituting
$\gamma'$ into
$K_{\rm M} = (1+6M_{\rm Pl}^2\gamma'^2)/\gamma^2$,
are:
\begin{align}
& K_{\rm M}^{(A)} = \frac{1}{\gamma^2}
  + \frac{3M_{\rm Pl}^2\,\alpha^2\sin^2(\phi/f)}
         {2f^2\,\gamma^4},
  \qquad & K_{\rm M}^{(B)}  = \frac{1}{\gamma^2}
  + \frac{3M_{\rm Pl}^2\,\alpha^2\cos^2(\phi/f)}
         {2f^2\,\gamma^4},&
  \\[6pt]
& K_{\rm M}^{(C)} = \frac{1}{\gamma^2}
  + \frac{3\alpha^2}{2\,\gamma^4},
  \qquad  & K_{\rm M}^{(D)} = \frac{1}{\gamma^2}
  + \frac{6\alpha^2\phi^2}
         {M_{\rm Pl}^2\,\gamma^4}. & 
\end{align}
The analytic slow-roll parameters, evaluated directly in $\phi$
via $\epsilon = \tfrac{M_{\rm Pl}^2}{2K}
\bigl(\tfrac{U_{,\phi}}{U}\bigr)^2$, are:
\begin{align}
\epsilon^{(A)} &=
  \frac{M_{\rm Pl}^2}{2f^2\,K}
  \;\frac{\sin^2\!\bigl(\tfrac{\phi}{2f}\bigr)\,
         \bigl[1-\alpha
           \bigl(1+\cos\tfrac{\phi}{f}\bigr)\bigr]^2}
        {\cos^2\!\bigl(\tfrac{\phi}{2f}\bigr)\,
         \bigl[1+\alpha
           \bigl(1+\cos\tfrac{\phi}{f}\bigr)\bigr]^2},
  \label{eq:epsA} \\[8pt]
\epsilon^{(B)} &=
  \frac{M_{\rm Pl}^2}{2f^2\,K}
  \;\frac{\bigl[\sin\tfrac{\phi}{f}
    \bigl(1+\alpha\sin\tfrac{\phi}{f}\bigr)
    +2\alpha\bigl(1+\cos\tfrac{\phi}{f}\bigr)
      \cos\tfrac{\phi}{f}\bigr]^2}
  {\bigl(1+\cos\tfrac{\phi}{f}\bigr)^2\,
   \bigl(1+\alpha\sin\tfrac{\phi}{f}\bigr)^2},
  \label{eq:epsB} 
\end{align}
\begin{align}  
\epsilon^{(C)} &=
  \frac{M_{\rm Pl}^2}{2\,K}
  \left[
    \frac{\tan\!\bigl(\tfrac{\phi}{2f}\bigr)}{f}
    +\frac{2\alpha}{M_{\rm Pl}\,\gamma^2}
  \right]^2,
  \label{eq:epsC} \\[8pt]
\epsilon^{(D)} &=
  \frac{M_{\rm Pl}^2}{2\,K}
  \left[
    \frac{\tan\!\bigl(\tfrac{\phi}{2f}\bigr)}{f}
    +\frac{4\alpha\,\phi}
          {M_{\rm Pl}^2\,\gamma^2}
  \right]^2.
  \label{eq:epsD}
\end{align}

In each expression, $K$ stands for $K_M$ or $K_P$ depending on
the formalism.  These are the formulas that are hardcoded in the
numerical implementation; no numerical differentiation of
$\gamma(\phi)$ is performed at any stage.

\subsection{Canonical field}
\label{app:canonical}

The canonical field $\chi(\phi)$, defined by
$d\chi/d\phi=\sqrt{K(\phi)}$, is integrated numerically via the
cumulative trapezoidal rule only for producing the Einstein-frame
potential profiles
(Figs.~\ref{fig:potential_comparisonA}--\ref{fig:potential_comparisonB}--\ref{fig:potential_comparisonC}--\ref{fig:potential_comparisonD}); the observables
themselves are computed directly in $\phi$ and do not require
$\chi$.  Where a closed-form expression is available (e.g.\
$\chi=\phi/\sqrt{1+2\alpha}$ for Coupling~A in the small-field
limit), it is used to cross-check the integration: the analytic
and numerical values of $\chi(\phi_*)$ agree to within
$\sim 0.3\%$ for $\alpha=0.5$, $f=1$.

\subsection{Validation}
\label{app:validation}

For the cosine coupling, the analytic expression for
$\epsilon(\phi)$ [Eq.~\eqref{eq:epsA}] is used to calibrate the
numerical solver.  The largest absolute deviation between the
analytic and numerical $\epsilon$ is of order $10^{-5}$, while
inside the inflationary window the typical difference is
$10^{-11}$--$10^{-19}$, confirming that the numerical scheme
reproduces the dynamics accurately.

As an additional check, the slow-roll parameters are computed in
two independent ways: (i)~from the analytic formula for each
coupling (Table~\ref{tab:couplings}), and (ii)~from centred
finite differences of $U(\phi)$ and $K(\phi)$ on the grid.  Both
routes yield consistent values of $(n_s,r)$ at the level of
$10^{-4}$.

\subsection{Physical domain and root-finding}
\label{app:domain}

The admissible parameter region is fixed by the positivity
condition $\gamma^2(\phi)>0$, which guarantees a positive effective
Planck mass.  For couplings with $\alpha<0$, this condition defines
a critical field value $\phi_{\rm crit}$ beyond which the effective
Planck mass becomes negative, and the upper limit of the grid is
set to $\phi_{\max}=\min(\phi_{\rm crit},\,f\pi)-\delta$
with $\delta=10^{-2}$, so that evaluations near the singular
boundary are avoided.

Two numerical subtleties are handled explicitly.
First, $\epsilon(\phi)=1$ may admit spurious roots outside the
slow-roll branch (near the potential maximum or where $K$ varies
rapidly); these are removed by restricting the root-finding
bracket to the first crossing from below on the inflationary
trajectory.
Second, parameter points for which the total number of $e$-folds
$N_{\max}<N_*$ are flagged as \texttt{NaN} and excluded from the
output, rather than forcing a solution that would lie outside the
physical domain.

The reported curves correspond to the parameter ranges
$3<f/M_{\rm Pl}<80$ for positive $\alpha$ and
coupling-specific ranges for negative $\alpha$, as quoted in the
figure captions.  For couplings with $\alpha<0$, the allowed values
of $f$ are restricted by requiring $f\pi<\phi_{\rm crit}$: if this
condition is violated, $\gamma^2$ would vanish before the field
reaches the minimum of the potential at $\phi=f\pi$.  Accordingly,
only values of $f$ satisfying $f\pi<\phi_{\rm crit}(\alpha)$ are
included in the scan.  In all cases, the scan uses $N_f=1000$
values of $f$ per $\alpha$ and is parallelised over $f$ using
\textsc{Joblib}.
\end{document}